\documentclass{ametsocV6.1}

\title{WRF model parameter calibration to improve the prediction of tropical cyclones over the Bay of Bengal using Machine Learning-based Multiobjective Optimization}

\authors{Harish Baki\aff{a} Sandeep Chinta\aff{a} C.Balaji\aff{a}\aff{b} Balaji Srinivasan\aff{a} \correspondingauthor{C.Balaji, balaji@iitm.ac.in}}

\affiliation{\aff{a}{Department of Mechanical Engineering, Indian Institute of Technology Madras, Chennai 600036, India} \\ \aff{b}{Center of Excellence in Atmospheric and Climate Sciences, Indian Institute of Technology Madras, Chennai 600036, India}}

\abstract{The prediction skill of a numerical model can be enhanced by calibrating the sensitive parameters that significantly influence the model forecast. The objective of the present study is to improve the prediction of surface wind speed and precipitation by calibrating the Weather Research and Forecasting (WRF) model parameters for the simulations of tropical cyclones over the Bay of Bengal region. Ten tropical cyclones across different intensity categories between 2011 and 2017 are selected for the calibration experiments. Eight sensitive model parameters are calibrated by minimizing the prediction error corresponding to 10m wind speed and precipitation,  using a multiobjective adaptive surrogate model-based optimization (MO-ASMO) framework. The 10m wind speed and precipitation simulated by the default and calibrated parameter values across different aspects are compared. The results show that the calibrated parameters improved the prediction of 10m wind speed by 17.62\% and precipitation by 8.20\% compared to the default parameters. The effect of calibrated parameters on other model output variables, such as cyclone track and intensities, and 500 hPa wind fields, is investigated. Eight tropical cyclones across different categories between 2011 and 2018 are selected to corroborate the performance of the calibrated parameter values for other cyclone events. Finally, the robustness of the calibrated parameters across different boundary conditions and grid resolutions is also examined. These results will have significant implications for improving the predictability of tropical cyclone characteristics. This allows us to better plan the adaptation and mitigation strategies and thus help in preventing the adverse effects on society.}

\begin{document}

\maketitle

%

\section{Introduction}
    Tropical cyclones (TCs) are considered one of the most devastating weather phenomena in tropical regions. The frequency of occurrence of severe cyclones in the Bay of Bengal (BoB) region is found to increase at an alarming rate of \(20\%\) per hundred years, and a two-fold increase has been observed during the intense cyclonic period of a year, i.e., May, October, and November \citep{singh2000changes}. A future projection of present global warming conditions shows that the intensity of the post-monsoon tropical cyclones over the BoB region will be much higher in the future compared to the present \citep{reddy2021impact}. Densely crowded coastal cities adjacent to the BoB, such as Chennai, Visakhapatnam, Bhubaneswar, and Kolkata, to name a few, have been affected by the widespread destruction caused by the tropical cyclones. An accurate prediction of the cyclone track and intensity during their landfall is inevitable to mitigate the destruction. \par
    The Weather Research and Forecasting (WRF) model is an atmospheric simulation system, which has been used by many scientists \citep{osuri2013real,carroll2021exploring} for research and operational purpose to monitor and predict the tropical cyclones across the globe. Similar to any other numerical model, the WRF model also suffers uncertainties that arise from the specification of initial and lateral boundary conditions, representation of model physics, and the specification of the model parameters \citep{di2015assessing}. Many researchers \citep{subramani2014new,chandrasekar2016,gogoi2021impact} have used data assimilation to improve the initial and lateral boundary conditions for the prediction of tropical cyclones over the BoB region. Several studies have been carried in the direction of parameterization schemes sensitivity as for example \cite{mukhopadhyay2011influence,pattanayak2012impact,osuri2012customization,chandrasekar2012,sandeep2018,Mohan2019} to study the role of physics schemes in accurately representing the model physics and their impact on the track, intensity, and rainfall prediction of tropical cyclones over the BoB region. However, very little research has been done on parameter optimization to improve the prediction of tropical cyclones over the BoB region in the WRF model.\par
    Parameter calibration is a process of adjusting the model parameters to match the model output to the observations. In recent years, several researchers \citep{yang2012some,wang2014evaluation,yang2015calibration,gong2016multiobjective,duan2017automatic,di2018assessing,di2019improving,di2020improving,chinta2020calibration} have attempted to calibrate the WRF model parameters to improve the model performance and recommended a two-step approach to be followed for the parameter calibration. They recommended that a sensitivity analysis be conducted to screen the most influential parameters. The WRF model consists of hundreds of tunable parameters and calibrating all of them requires tremendous computational power. In addition, calibrating the parameters which show no influence on the fundamental meteorological variables may lead no improvement in the model accuracy. Thus, there is a need to identify the most sensitive parameters which greatly influence the fundamental meteorological variables. Several researchers have evaluated the sensitivity of the WRF model parameters using qualitative and quantitative methods \citep{wang2013parameter,green2014sensitivity, quan2016evaluation,di2017parametric,houle2017exploring,ji2018assessing,wang2020assessing,sandeep2020assessment,baki2021sensitivity}. \par
    Once the sensitive parameters are estimated, the second step is to calibrate these parameters with respect to the fundamental meteorological variables. \cite{yang2012some} calibrated five parameters of the Kain-Fritsch cumulus schemes of the WRF model using Multiple Very Fast Simulated Annealing (MVFSA) \citep{jackson2008error} algorithm and reported that the precipitation bias is reduced by a considerable amount. The authors also applied a similar methodology to calibrate five parameters of the Kain-Fritsch cumulus scheme for the predictions of East Asian Summer Monsoon and reported an improvement in precipitation and surface energy features with the calibrated parameters \citep{yang2015calibration}. \cite{wang2014evaluation} developed the Adaptive Surrogate Modelling Based Optimization (ASMO) methodology that has been used by multiple researchers to calibrate the model parameters. \cite{gong2014multi,gong2016multiobjective} extended the ASMO method for the optimization of multiple objective functions as MultiObjective - Adaptive Surrogate Modelling based Optimization (MO-ASMO) and calibrated the parameters of the land surface model. \cite{duan2017automatic} adopted the ASMO method to calibrate the WRF model parameters to improve the prediction of summer monsoon over the Greater Beijing Area. With a promise found using the ASMO method, \cite{di2018assessing,di2019improving,di2020improving} conducted a series of studies to calibrate the WRF model parameters across different weather phenomena over various parts of China to improve the prediction of different meteorological variables. 
    \cite{chinta2020calibration} adopted the MO-ASMO method to calibrate the WRF model parameters and reported that an improvement is observed for the predictions of precipitation, surface air pressure, surface air temperature, and wind speed at 10m, during the Indian Summer Monsoon (ISM).\par
    However, the optimal parameters that are obtained from a calibration study will vary with the output variable, the type of simulation event, and the geographical location. A review of the pertinent literature shows that  very little research has been done on the WRF model parameter calibration to improve the prediction of tropical cyclones over the Bay of Bengal region. In view of this, the present study focuses on improving the prediction of tropical cyclones across different intensity categories by employing parameter calibration. The model parameters utilized in the present study are obtained based on a previous investigation conducted by \cite{baki2021sensitivity}, where three different sensitivity analysis methods, namely Morris One-At-a-Time (MOAT), Multivariate Adaptive Regression Splines (MARS), and Gaussian Process Regression (GPR) based Sobol' were applied to screen the most influential parameters for the simulations of tropical cyclones over the Bay of Bengal region. The selected parameters are tuned in this study by minimizing the error between the model simulations and the observations of precipitation and 2m wind speed for the simulations of ten tropical cyclones by employing the MO-ASMO method. The performance of the calibrated parameter values is evaluated by simulating another eight tropical cyclones over the same region. In addition, the robustness of the calibrated parameters across different spatial resolutions and driving data is also performed. \par
    This paper is organized as follows. Section 2 describes the methodology of MO-ASMO that is adopted in the present study. Section 3 presents the WRF model configuration, selected parameters, simulation events, data, and calibration setup. Section 4 lists the significant results obtained from the calibration and validation experiments. Conclusions from this study are presented in Section 5.
\section{MO-ASMO Method}\label{MO-ASMO}
    Compared to standard parameter optimization approaches, the MultiObjective Adaptive Surrogate Model-based Optimization (MO-ASMO) technique is particularly well suited for parameter optimization of large, complicated dynamic models such as the WRF model with high computing demands. Because the MO-ASMO technique relies mostly on the statistical surrogate model rather than the real physical model to find the ideal model solution (typically the least error between simulation and observation), the optimization convergence speed will be substantially enhanced. The MO-ASMO process consists of four steps: (1) obtaining initial sampling of the physical model, (2) construction of surrogate models for each variable using the initial samples, (3) generation of new adaptive samples based on optimization of the individual surrogate models, (4) generating new adaptive samples based on multiobjective optimization, and (5) obtaining final optimized parameter values using multiple-criteria decision analysis. The following are detailed explanations of these stages. \par
    \begin{enumerate}
        \item A robust surrogate model requires a substantial amount of input and target data from real model simulations. As a result, the real model is simulated a sufficient number of times with various parameter values within their defined ranges. An appropriate sampling approach is used to obtain the sets of parameter values needed to run the real model, which will consistently distribute the samples throughout the parameter space. In the present study, the Quasi-Monte Carlo \citep{sobol1967distribution} sampling design is used for parameter sampling because of its capacity to traverse the whole parameter space with relatively lean sampling points and have been used by many researchers for this purpose \citep{di2018assessing,di2020improving}. The initial parameter sample sets are generated using the QMC sampling design. The numerical simulations of the actual model are performed with these parameter values. Following this, the simulation errors are computed by comparing the model simulations with the observations. This way, the initial samples of the physical model are obtained, which consists of the parameter values as inputs and their accompanying simulations errors as the targets.
        \item Construct surrogate models for individual model output variables using the initial samples to represent the actual physical model. The surrogate models are constructed using statistical regression approaches. For example, Gaussian Process Regression (GPR), Support Vector Machine (SVM) Multivariate Adaptive Regression Spline (MARS), Artificial Neural Network (ANN), and the sum of trees (SOT) are a few of the regression methods available. The GPR has been used by many researchers \citep{chinta2020calibration,di2018assessing} and has been found to be efficient compared to other regression methods, and in view of this, it has been employed to construct the surrogate models in the present study.
        \item Upon constructing the GPR surrogate models, a rapid optimization method is utilized to optimize the hyperparameters of the surrogate models. The Shuffled Complex Evolution \citep{duan1993shuffled} and the Genetic Algorithm (GA) \citep{schmitt2001theory}, to name a few, are some of the rapid optimization algorithms available. The GA is an adaptive heuristic exploration method that uses knowledge from the previous iteration to intelligently exploit random search and lead the exploration towards the optimum solution field, which has been used in this study. With the newly obtained parameter values using the GA, the original physical model is run, and the model simulation errors are evaluated. These parameter values and simulation errors are given as new data points to the surrogate models to improve their accuracy. This way, the GPR models continue to modify as the new data points are generated. This process is called adaptive sampling, which is continued till a predefined convergence criterion for the parameter optimization is met.
        \item Once optimized surrogate models for the individual variables are constructed, multiobjective optimization is performed to obtain final optimized parameters, which minimize the model simulation errors of all the variables at once. For this purpose, the non-dominated sorting genetic algorithm II (NSGA-II) developed by \cite{deb2002fast} is employed in this study. The NSGA-II reaches the defined objective in six steps, that are (i) population initialization, (ii) non-dominated sorting of the population, (iii) assigning the crowding distance values to the fronts, (iv) selection of individuals, (v) genetic operator to generate new offspring, (vi) and combining new offspring and the current generation. The NSGA-II algorithm generates a Pareto front with optimum solutions called non-dominated solutions. A non-dominated solution is one that achieves a good balance between all goals without degrading any of them. Similar to the previous adaptive sampling, 25\% of the solutions from the Pareto front that have the highest crowding distance values are chosen for the numerical simulations using the actual model, and the simulations errors are evaluated. The surrogate models are updated with the newly generated samples. This stage is continued until it reaches convergence, which is when the number of simulations of the physical model hits a predefined value.
        \item NSGA-II produces a final Pareto front that frequently comprises hundreds of solutions, and selecting one final optimal solution out of these requires a multi-criteria decision analysis. For this purpose, the Technique for Order of Preference by Similarity to Ideal Solution (TOPSIS) is adopted in this study. The individual variable objectives are multiplied by equal weights, and are used to find the positive and negative ideal solutions. The final optimal solution has the shortest distance from the positive ideal solution and the longest distance from the negative ideal solution. 
    \end{enumerate}
\section{Design of numerical experiments}
\subsection{WRF model configuration and parameters selected for calibration}
\begin{figure}
    \centering
    \includegraphics[width=0.4\textwidth,angle=90,origin=c]{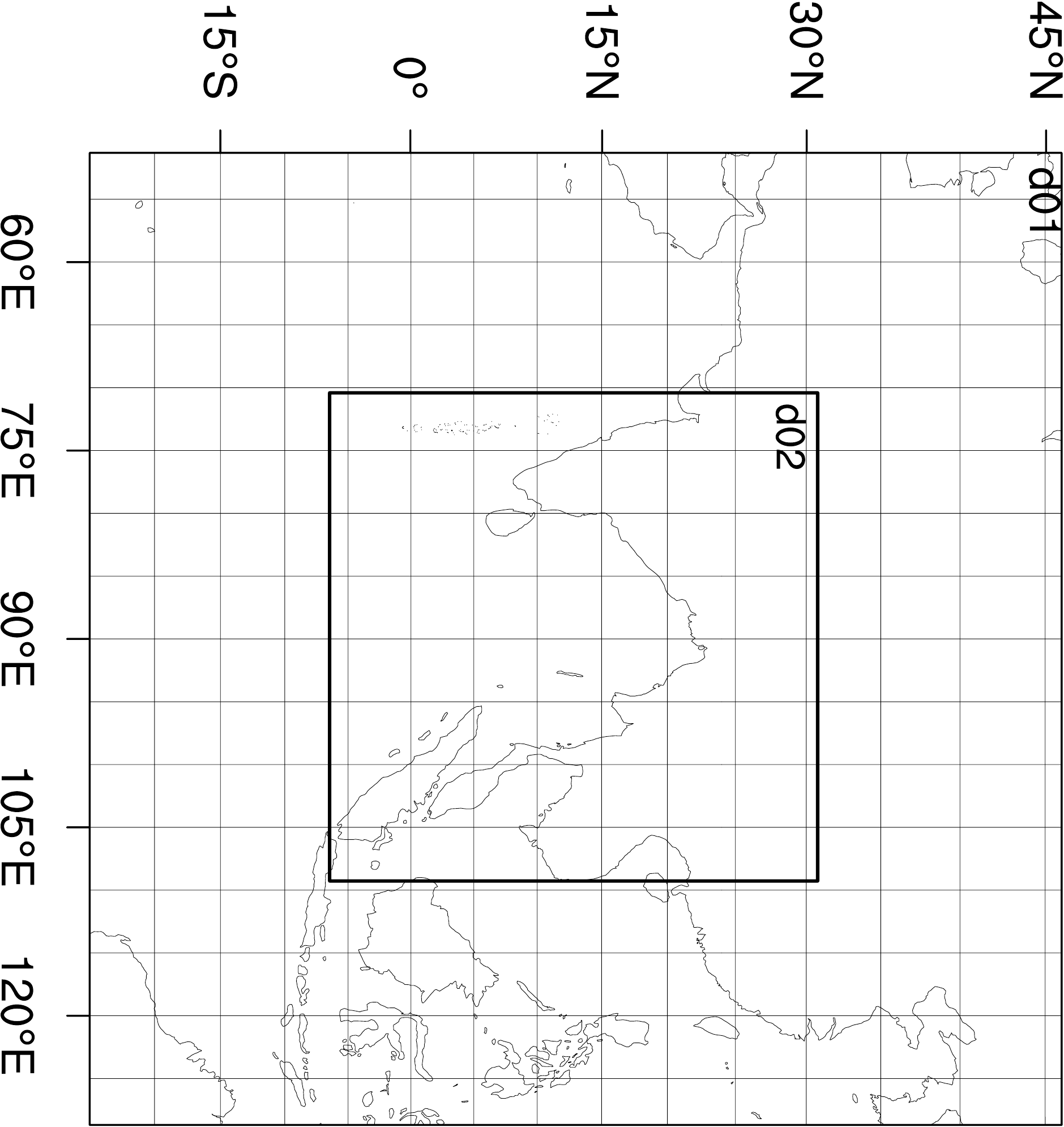}
    \caption{An illustration of the WRF model domain configuration. The parent domain (d01) consists \(240 \times 240\) grid points with 36km resolution and the nested domain (d02) consists of \(360 \times 360\) grid points with 12km resolution.}
    \label{domains}
    \end{figure}

    The WRF model version 3.9.1 \citep{skamarock2008description} is used for parameter calibration to improve the model prediction of tropical cyclones over the Bay of Bengal. The two domains considered in this study are illustrated in Figure \ref{domains}. The inner domain (d02) covers the entire Bay of Bengal and the adjacent Indian coastal regions. The parent domain (d01) consists of \(240 \times 240\) grid points with a spatial resolution of 36 kilometers, and the inner domain (d02) consists of \(360 \times 360\) grid points with a spatial resolution of 12 kilometers. The domain is vertically divided into terrain-following coordinates of 50 sigma layers with a fine resolution below the boundary layer. The resolution gradually becomes coarse and reaches a height of 50 hPa. The initial and boundary conditions for the simulations are taken from the NCEP FNL (National Centers for Environmental Predictions) operational global analysis and forecast data at \(1^{\circ}\times 1^{\circ}\) resolution with a six-hourly interval \citep{cisl_rda_ds083.2}. The model domains d01 and d02 are integrated with time-steps of 90 s and 30 s, respectively. \par
    The WRF model represents the unresolved physical processes through seven parameterization schemes: surface layer physics, land surface physics, shortwave radiation, longwave radiation, planetary boundary layer physics, microphysics, and cumulus physics. The parameterization schemes used in the present work are selected based on the studies of \cite{baki2021sensitivity}, which are: MM5 similarity scheme \citep{beljaars1995parametrization} for surface layer physics, Unified Noah land surface model \citep{mukul2004implementation} for land surface physics, Dudhia shortwave scheme \citep{dudhia1989numerical} for shortwave radiation, rapid radiative transfer model \citep{mlawer1997radiative} for longwave radiation, WRF Single-Moment 6-class (WSM6) scheme \citep{hong2006wrf} for microphysics, Kain-Fritsch \citep{kain2004kain} for cumulus physics, and Yonsei University Scheme (YSU) \citep{hong2006new} for planetary boundary layer physics. The parameterization schemes contain a vast number of tunable parameters, and calibrating all of them is an impossible task. \cite{baki2021sensitivity} conducted a sensitivity study of 24 parameters from the same physics schemes for tropical cyclones over the Bay of Bengal and reported that a total of eight parameters were found to be most sensitive. Thus, these parameters are selected for calibration to improve the prediction of tropical cyclones over the BoB. The selected parameters and their corresponding ranges are listed in Table \ref{Adjustable_parameters}.
    \begin{table}
    \caption{List of the sensitive parameters from seven physics schemes of the WRF model.}
    \scriptsize
    \centering
    \renewcommand\arraystretch{1.5}
    \begin{tabular}{p{0.2\linewidth}p{0.11\linewidth}p{0.08\linewidth}p{0.11\linewidth}p{0.35\linewidth}}
        \hline
        \textbf{Scheme}    & \textbf{Parameter} & \textbf{Default} & \textbf{Range} & \textbf{Description} \\
        \hline
        Surface layer & znt\_zf &	1 &	[0.5 2] &	Scaling related to surface roughness \\
        	& karman &	0.4 &	[0.35 0.42] &	Von Kármán constant \\
        \hline
         Cumulus &   pe &	1 &	[0.5 2] &	The multiplier for entrainment mass flux rate \\
        \hline
         Microphysics &	ice\_stokes\_fac &	14900 &	[8000 30000] &	Scaling factor applied to ice fall velocity (s-1) \\
        \hline
           Shortwave radiation &	cssca\_fac &	1.00E-05 &	[5E-6 2E-5] &	Scattering tuning parameter (m2 kg-1) \\
        \hline
         Longwave &	Secang &	1.66 &	[1.55 1.75] &	Diffusivity angle for cloud optical depth computation \\
        \hline
         Land surface	&    porsl &	1 &	[0.5 2] &	The multiplier for the saturated soil water content \\
        \hline
         Planetary boundary layer  &	pfac &	2 &	[1 3] &	Profile shape exponent for calculating the momentum diffusivity coefficient \\
        \hline
    \end{tabular}
    \label{Adjustable_parameters}
    \end{table}
\subsection{Simulation events and Data}
\begin{figure}
    \centering
    \includegraphics[width=0.75\linewidth]{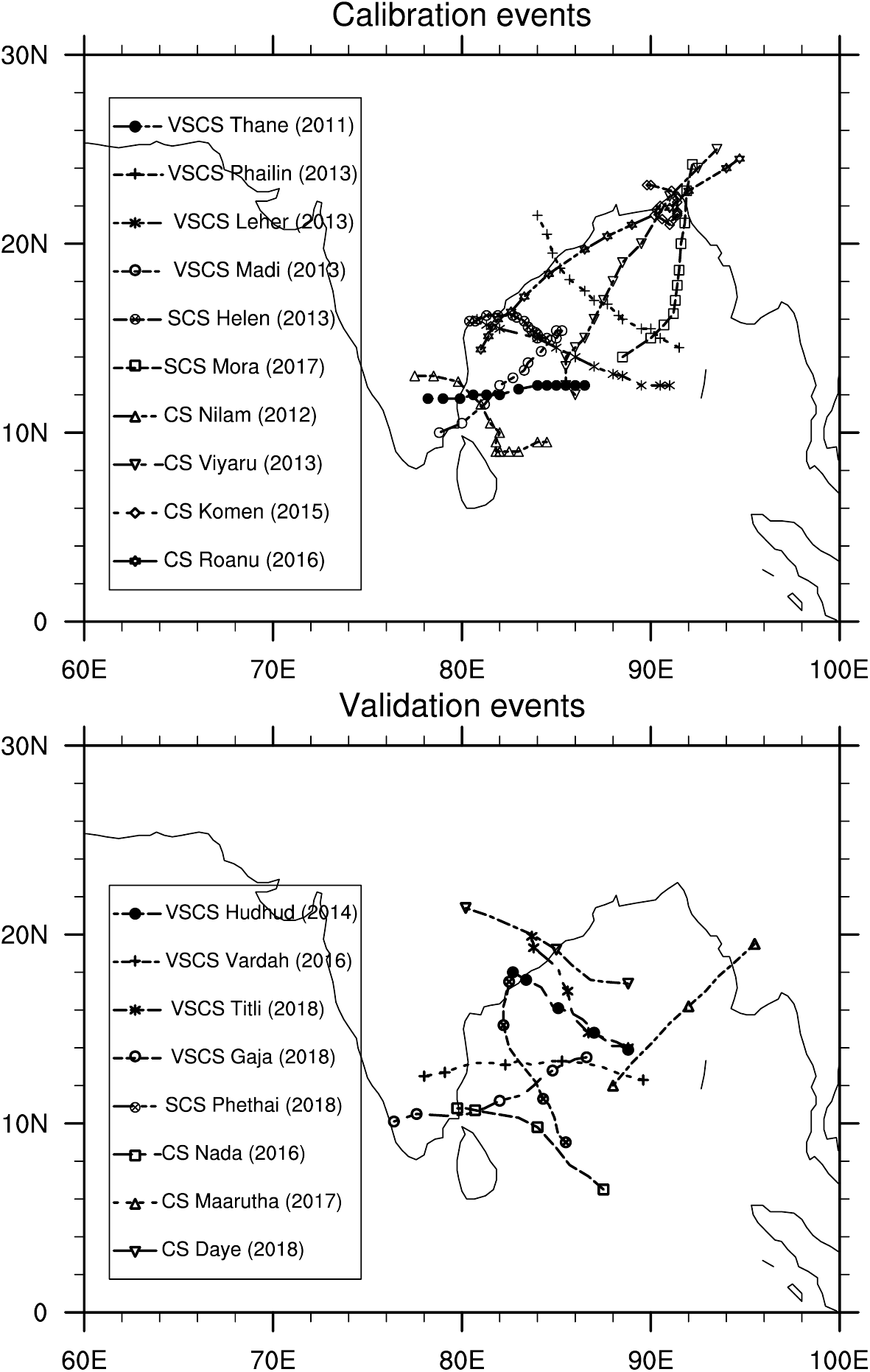}
    \caption{An illustration of the IMD observed tracks of tropical cyclones selected for the calibration and validation experiments.}
    \label{IMD_tracks}
    \end{figure}

    In this study, eighteen tropical cyclones that formed in the Bay of Bengal between 2011 and 2017 were chosen for the numerical experiments. The cyclones chosen are from different intensity categories that are categorized by the Indian Meteorological Department (IMD) according to the Maximum Sustained Surface Wind Speed (MSW). The tropical cyclone categories used in this study are Very Severe Cyclonic Storm (64-119 knots), Severe Cyclonic Storm (48-63 knots), and Cyclonic Storm (34-47 knots) \citep{srikanth2012study}. A total of 18 tropical cyclones are chosen for this study, and their details such as category, landfall time, and simulation duration are listed in Table \ref{cyclones}, and the corresponding IMD observed tracks are illustrated in Figure \ref{IMD_tracks}. Ten events are selected from the 18 cyclones for the optimization experiments to obtain the parameters that minimize the model simulation errors. These events are termed as calibration events hereafter. The remaining eight events are selected to validate the obtained optimized parameter values. These events are termed validation events hereafter. The cyclones are simulated for 96 hours, including 12 hours of spin-off period, 72 hours before landfall, and 12 hours following landfall.\par
    The calibration of parameters is conducted to minimize the simulation errors of wind speed at 10 meters above the ground (10m wind speed) and the total precipitation of all the events. The model simulated 10m wind speed data is compared with the Indian Monsoon Data Assimilation and Analysis (IMDAA) data \citep{ashrit2020imdaa}, which is available at \(0.12^{\circ} \times 0.12^{\circ}\) resolution, and the model simulated precipitation data is compared with the Integrated Multi-satellitE Retrievals for GPM (IMERG) dataset \citep{huffman2017gpm}, which is available at \(0.1^{\circ} \times 0.1^{\circ}\) resolution. In addition to these datasets, the IMD observations of central sea level pressure (CSLP), MSW, and cyclone track are also used to evaluate the model simulations. The WRF model variables are stored at 6-hour intervals. In view of this, the observations are also taken at this latency for comparison.
    \begin{table}
    \caption{Overview of the tropical cyclones selected for the calibration and validation experiments.}
    \scriptsize
    \centering
    \renewcommand\arraystretch{1.5}
    \begin{tabular}{p{0.06\linewidth}p{0.15\linewidth}p{0.3\linewidth}p{0.38\linewidth}}
    \hline
    \textbf{Index}  & \textbf{Cyclone}    & \textbf{Landfall time} & \textbf{Simulation duration} \\
    \hline
    \hline
        &       & Calibration events  &   \\
    \hline
    A   &   VSCS Thane      & 0100 - 0200 UTC 30th Dec 2011 & 2011-12-26\_18:00:00 to 2011-12-30\_18:00:00 \\
    B   &   VSCS Phailin    & 1700 UTC 12th Oct 2013        & 2013-10-09\_06:00:00 to 2013-10-13\_06:00:00 \\
    C   &   VSCS Leher      & 0830 UTC 28th Nov 2013        & 2013-11-25\_00:00:00 to 2013-11-29\_00:00:00 \\
    D   &   VSCS Madi       & 1700 UTC 12th	Dec 2013        & 2013-12-09\_06:00:00 to 2013-12-13\_06:00:00 \\
    E   &   SCS Helen       & 0800 - 0900 UTC 22nd Nov 2013 & 2013-11-19\_00:00:00 to 2013-11-23\_00:00:00 \\
    F   &   SCS Mora        & 0400 - 0500 UTC 30th may 2017 & 2017-05-26\_18:00:00 to 2017-05-30\_18:00:00 \\
    G   &   CS Nilam        & 1030 - 1100 UTC 31st Oct 2012  & 2012-10-28\_00:00:00 to 2012-11-02\_00:00:00 \\
    H   &   CS Viyaru       & 0230 UTC 16th May 2013        & 2013-05-12\_18:00:00 to 2013-05-16\_18:00:00 \\ 
    I   &   CS Komen        & 1400 - 1500 UTC 30th July 2015 & 2015-07-27\_06:00:00 to 2015-07-31\_06:00:00 \\
    J   &   CS Roanu        & 1000 UTC 21st May 2016        & 2016-05-18\_00:00:00 to 2016-05-22\_00:00:00 \\
    \hline
        &       & Validation events  &   \\
    \hline
    K	&	VSCS Hudhud	&	0630 UTC 12th Oct 2014	&	2014-10-08\_12:00:00 to 2014-10-12\_12:00:00	\\
    L	&	VSCS Vardah	&	0930 – 1030 UTC 12th Dec 2016	&	2016-12-09\_00:00:00 to 2016-12-13\_00:00:00	\\
    M	&	VSCS Titli	&	0000 UTC 11th Oct 2018	&	2018-10-07\_12:00:00 to 2018-10-11\_12:00:00	\\
    N	&	VSCS Gaja	&	1900 – 2000 UTC 15th Nov 2018	&	2018-11-12\_12:00:00 to 2018-11-16\_12:00:00	\\
    O	&	SCS Phethai	&	1400 – 1500 UTC 17th Dec 2018	&	2018-12-14\_06:00:00 to 2018-12-18\_06:00:00	\\
    P	&	CS Nada	&	2230 – 2330 UTC 1st Dec 2016	&	2016-11-28\_12:00:00 to 2016-12-02\_12:00:00	\\
    Q	&	CS Maarutha	&	1800 – 1900 UTC 16th April 2017	&	2017-04-13\_12:00:00 to 2017-04-17\_12:00:00	\\
    R	&	CS Daye	&	1900 – 2000 UTC 20th Sept 2018	&	2018-09-17\_12:00:00 to 2018-09-21\_12:00:00	\\
    \hline
    \end{tabular}
    \label{cyclones}
    \end{table}
    
\subsection{Experimental setup for parameter calibration}
    In the present study, the MO-ASMO method \citep{chinta2020calibration} is used to optimize the eight sensitive parameters of the WRF model with respect to the surface wind speed and precipitation simulations. The MO-ASMO is achieved in five steps. First, the initial parameter sets are acquired by sampling the parameters according to their ranges and distributions. Studies of \cite{di2018assessing,di2020improving} suggest that the Quasi-Monte Carlo (QMC) sampling method is suitable to produce uniform samples at higher dimensions. Thus, the QMC Sobol' sampling design \citep{sobol1967distribution}, which is developed in Uncertainty Quantification Python Laboratory (UQ-PyL) \citep{uqpyl} is employed to generate the initial parameter samples sets. \citep{di2018assessing} stated that ten times the dimensionality of the parameters is sufficient for the initial samples. Hence, 100 sample sets of 8 parameters are generated using the QMC Sobol' sampling design. With the obtained parameter sample sets, the WRF model simulations are performed for the calibration events that are listed in Table \ref{cyclones} and illustrated in Figure \ref{IMD_tracks}(a). The WRF model simulations of 10m wind speed and precipitation are compared with the observations to calculate the root mean square error (RMSE) of each variable.
    \begin{equation}\label{eq_RMSE}
        RMSE = \sqrt{\frac{\Big[\sum_{n=1}^N\sum_{t=1}^T \sum_{i=1}^I(sim_{it}^n-obs_{it}^n)^2\Big]}{I\times T\times N}}
    \end{equation}
    Where \textit{sim} is the simulated quantity, \textit{obs} is the observed quantity, \textit{I} is the total number of horizontal grid points in the d02 domain, \textit{T} is the dimension of simulation duration at 6 hours interval, and \textit{N} is the total number of calibration events. In addition to these simulations with 100 parameter sample sets, the WRF simulations with default parameters are also conducted and the corresponding RMSE values are evaluated. A normalized RMSE value for each of the 100 initial samples is evaluated by dividing with the default RMSE value. 
    \begin{equation}\label{eq_NRMSE}
        NRMSE_i = \frac{RMSE_i}{RMSE_{def}}
    \end{equation}
    Where \(RMSE_i\) is the RMSE value of the particular parameter sample and the \(RMSE_{def}\) is the default RMSE value. Finally, two datasets of \(100 \times 8\) containing 100 samples sets with 8 parameter values as inputs and corresponding outputs of \(100 \times 2\) containing 100 NRMSE values of 10m wind speed and precipitation are obtained. Second, using these input and output datasets, two GPR surrogate models are constructed for each variable. Therefore, the GPR model of 10m wind speed takes eight inputs of parameter values and one output of 10m wind speed NRMSE value. Similarly, the GPR model of precipitation takes the same inputs but one output of precipitation NRMSE value. \par
    The third step following the construction of the surrogate models using the initial samples is the use of the genetic algorithm to determine the optimal parameter values. With these new parameter values, the WRF model simulations are performed, and corresponding NRMSE values are evaluated. These new parameter values and corresponding NRMSE values are added to the initial dataset, and the existing surrogate models are updated. This process of adaptive parameter sampling and updating of the surrogate models is repeated until the convergence criteria for parameter optimization is met. When the NRMSE value does not decrease continuously for ten iterations, convergence is declared. The parameter values obtained after the convergence of each surrogate model can be considered as the optimized parameters for that particular variable. This way, one set of optimized parameter values are obtained for (i) 10m wind speed and (ii) precipitation. However, the primary goal of the current study is to calibrate the WRF model parameter values to minimize the NRMSE values of 10m wind speed and precipitation simulations together. Thus, multiobjective optimization is performed with the help of NSGA-II in the fourth step.\par
    The NSGA-II algorithm, developed in python language by \cite{pymoo} is utilized in the present study. The updated surrogate models of 10m wind speed and precipitation and the ranges of parameter values are provided to the NSGA-II algorithm. The algorithm searches for optimized values with a population of 80 for 200 generations and outputs a final optimized values of \(80 \times 8\) matrix. Out of these 80 solutions, (\(25\%\) of \(80=\)) 20 solutions with the highest crowding distances are selected for the numerical simulations, and the simulation errors along with the input parameters are used to update the existing surrogate models. This process is repeated for five times which results in (\(20 \times 5 = \)) 100 model simulations that reaches the targeted number of model runs limit \citep{chinta2020calibration}. The final output consists of nondominated solutions on the Pareto front, out of which the best solution is identified with the application of TOPSIS multi-decision criteria in the final step. \par
    The NRMSE value is used to evaluate the accuracy of the WRF model simulations, with \(NRMSE < 1\) implying a better simulation compared to the default, and the parameter values with the least NRMSE value are said to be the optimal parameter values. Taylor statistics presented on a Taylor diagram \citep{taylor2001summarizing} are another metric used to assess the accuracy of the model simulations. The Taylor statistics consists of a correlation coefficient, centered pattern RMS error, and normalized standard deviation. These statistics can be plotted on a single Taylor diagram, which is very helpful to explain the differences between the performances of multiple models and multiple variables. 
    Apart from the NRMSE and Taylor statistics, another quantitative metric used to evaluate the performance of the default and calibrated parameters for the simulations precipitation is Structure, Amplitude, and Location (SAL) indices proposed by \cite{wernli2008sal}. The precipitation fields are segregated into objects based on the thresholds, and the S, A, and L calculations are evaluated. A precipitation forecast with S, A, and L values close to zero signifies a higher accuracy of that simulation. \par
\section{Results and discussion}
\subsection{Parameter calibration results}
\subsubsection{Convergence of single and multiobjective optimization runs}
\begin{figure}
    \centering
    \includegraphics[width=0.6\linewidth]{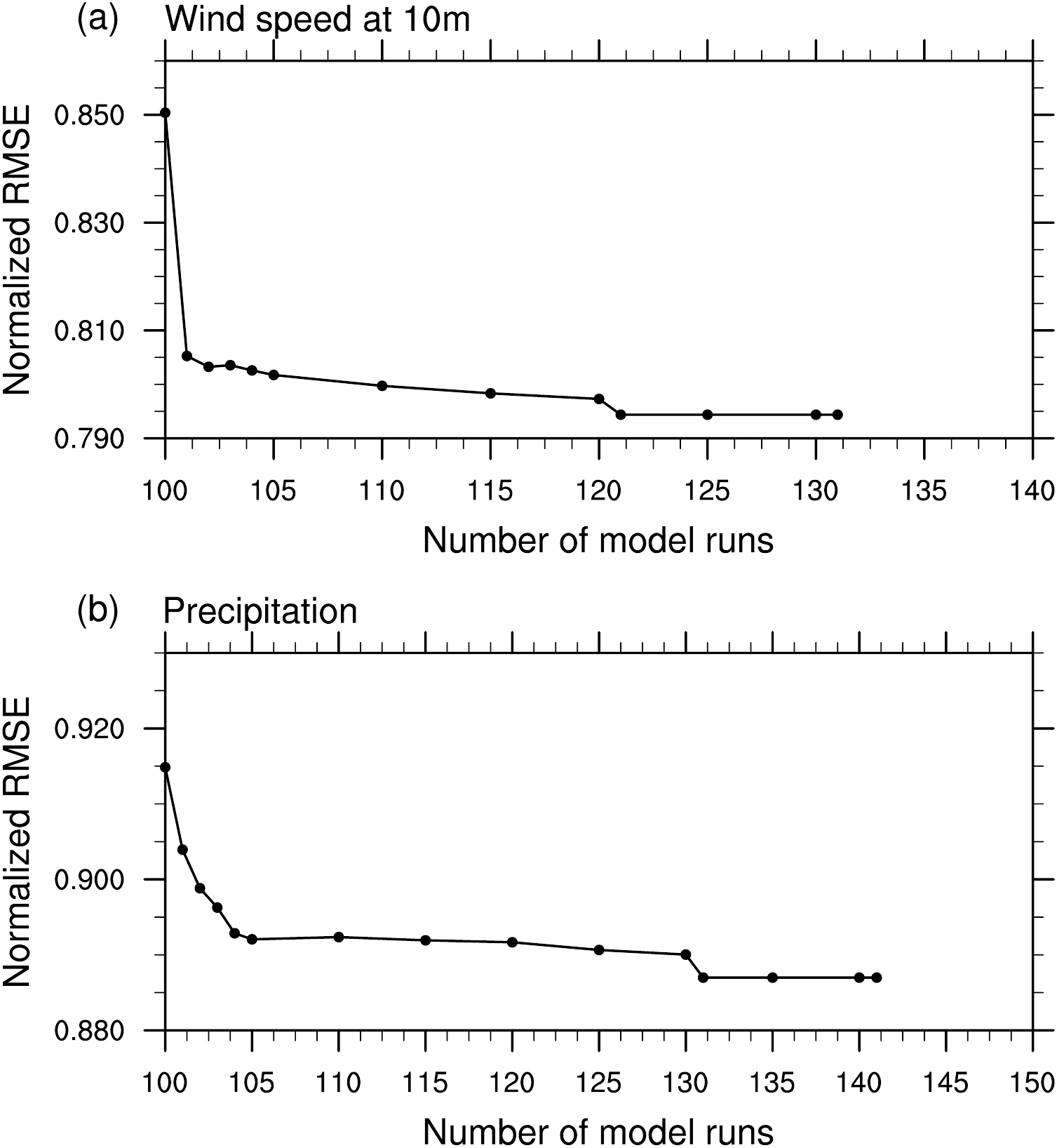}
    \caption{Convergence trends of the surrogate models for the optimization of (a) 10m wind speed and (b) precipitation.}
    \label{convergence_trend}
\end{figure}

\begin{figure}
    \centering
    \includegraphics[width=0.6\linewidth]{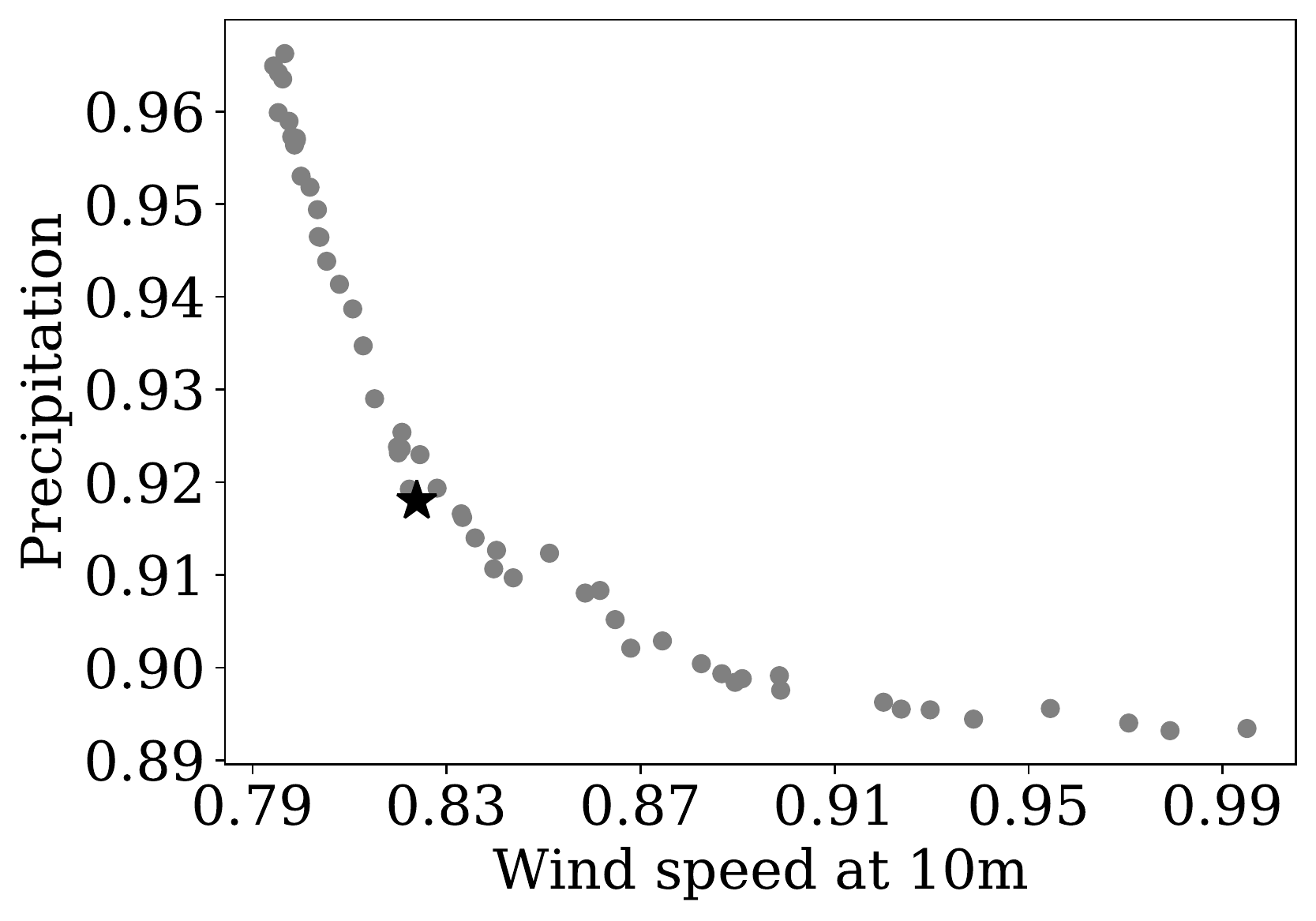}
    \caption{Pareto front of the non-dominated solutions obtained from the NSGA-II multiobjective optimization with the black star indicating the final solution obtained by aplying the TOPSIS method.}
    \label{paritofront}
\end{figure}

Figure \ref{convergence_trend} presents the convergence trends of the parameter optimization loops for the two quantities of interest, namely (i) 10m wind speed and (ii) precipitation, during the 84 hours simulations of the calibration events together. The minimum NRMSE values from the initial m samples are found to be 0.850 for the 10m wind speed and 0.915 for the precipitation, which implies that the simulation errors are decreased by 15\% (10m wind speed) and 9.05\% (precipitation) using the initial 100 samples itself when compared with the default parameters. The additional number of model runs required for the convergence of 10m wind speed and precipitation optimization are 21 and 31, respectively. The parameter values obtained at these iterations are termed as the corresponding variable optimized parameters. Beyond these runs, no reduction in the NRMSE is observed for the next ten iterations. The minimum NRMSE values found at the converged iterations are listed in Table \ref{min_cal_NRMSE}, show that the simulation errors are reduced by 20.56\% (10m wind speed) and 11.30\% (precipitation). Since the present study aims to improve the prediction of 10m wind speed and precipitation together, the NSGA-II optimization is employed to perform multiobjective optimization. The NSGA-II produces a Pareto front of non-dominated solutions after five iterations as shown in Figure \ref{paritofront}, in which the abscissa indicates the NRMSE values of 10m wind speed and the ordinate indicates the NRMSE values of the precipitation. The NRMSE values of 10m wind speed range from 0.79 to 0.99, and that of the precipitation range from 0.89 to 0.96. The Pareto front clearly shows that the least NRMSE value of 10m wind speed has the highest NRMSE value of precipitation, which indicates that improving one variable will degrade another variable. Thus, the TOPSIS method with a weight of 0.5 to each of the two objectives is employed to obtain the final best solution, which is shown as a black star in Figure \ref{paritofront}. Hereafter the parameter values in the best solution are considered as the calibrated parameter values. A total of 252 (100+21+31+100) WRF model runs were required to obtain the final calibrated parameter values. The final NRMSE values after the calibration, as shown in Table \ref{min_cal_NRMSE} reveal that a considerable deviation is observed from the single objective optimized NRMSE values. This can be explained by Figure \ref{paritofront}, which shows that improving one variable will lead to the degradation in another variable. Finally, the simulation errors after the calibration are reduced by 17.62\% (10m wind speed) and 8.20\% (precipitation).

\begin{table}
    \caption{Final NRMSE values obtained from the optimization of 10m wind speed, precipitation, and the combination (multiobjective) of 10m wind speed and precipitation.}
    \scriptsize
    \centering
    \renewcommand\arraystretch{1.75}
    \begin{tabular}{p{0.3\linewidth}p{0.15\linewidth}p{0.15\linewidth}}
    \hline
    \textbf{Optimization runs}    & \textbf{10m wind speed} & \textbf{precipitation} \\
    \hline
    \hline
    10m wind speed optimization   & \textbf{0.7944} &  0.9637\\
    precipitation optimization  & 1.1853 & \textbf{0.8870} \\
    multiobjective optimization     & 0.8239 & 0.9180 \\
    \hline
    \end{tabular}
    \label{min_cal_NRMSE}
\end{table}

\subsubsection{Comparison of RMSE values for the WRF simulations with default and optimized parameters}
\begin{figure}
    \centering
    \includegraphics[width=\linewidth]{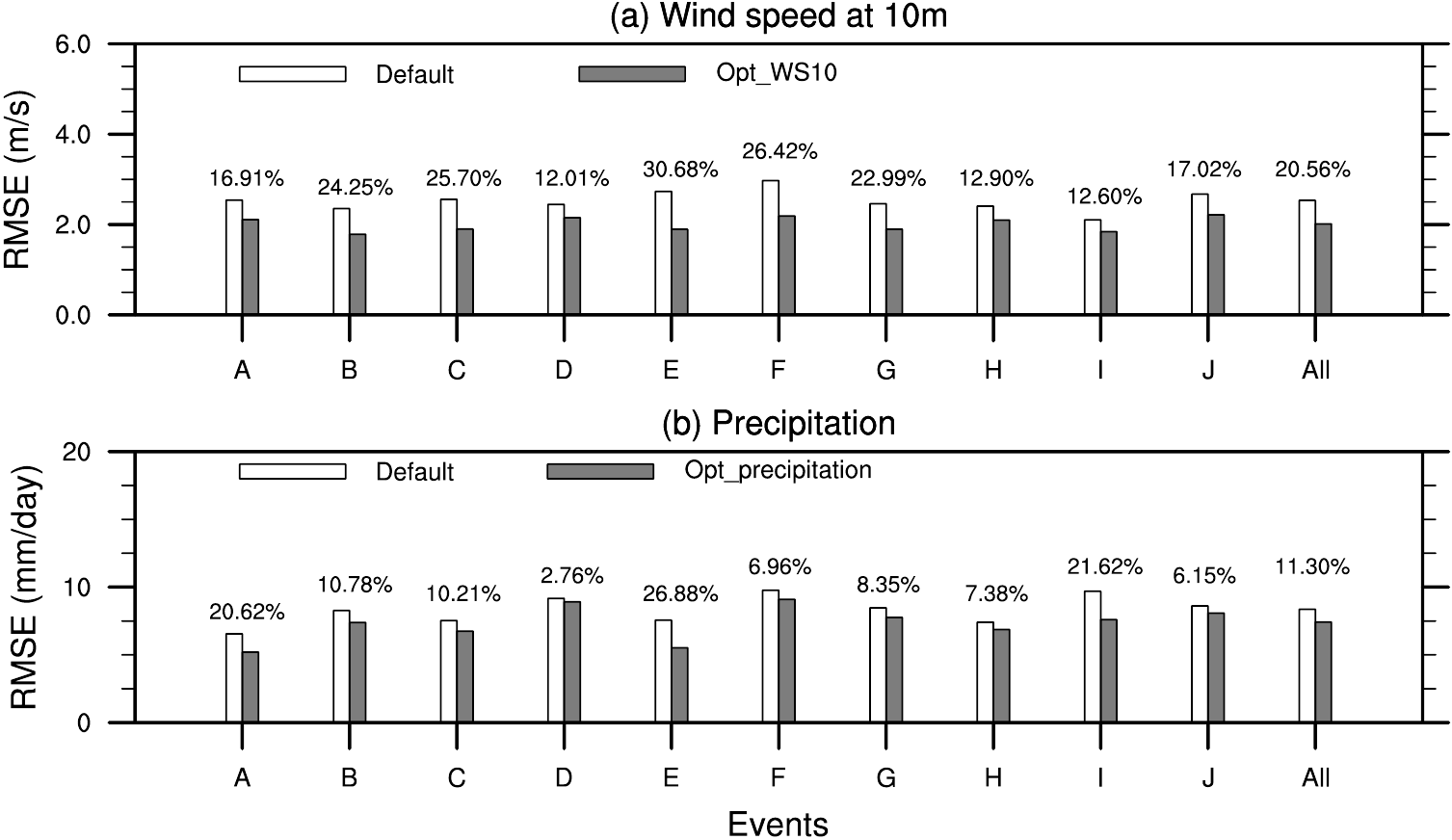}
    \caption{RMSE values of the (a) 10m wind speed simulations using the parameters obtained from the 10m wind speed optimization and (b) precipitation simulations using the parameters obtained from the precipitation optimization, in comparison with the default RMSE values, for the calibration events (A)–(J). The percentage values indicate the reduction in the RMSE value of variables when using the optimized parameters compared to the default parameters.}
    \label{single_objective}
\end{figure}

Figure \ref{single_objective} presents the RMSE values of 10m wind speed simulated with the optimized parameters of 10m wind speed and precipitation simulated with the optimized parameters of precipitation, in comparison with the default parameter simulations, for the ten calibration events (A-J). The results clearly show a significant reduction in the simulation RMSE values by the optimized parameters compared to the default parameters. The reduction in errors with the predictions of 10m wind speed is the highest for event E (30.68\%) and the lowest for event D (12.01\%). As stated earlier, the improvement in 10m wind speed for all the events combined together is 20.56\%. Similarly, the precipitation simulation RMSE reduction ranges from 2.76\% for event D to 26.88\% for event E, and the combined reduction is 11.30\% for all the events together. The results show that the individual optimization of 10m wind speed or precipitation can significantly improve the respective variable simulations. The parameters obtained from the individual optimizations can be adopted in any study to improve the respective outcomes of numerical simulations of tropical cyclones over the Bay of Bengal region, i.e. when only that particular variable is of interest to us. \par
\begin{figure}
    \centering
    \includegraphics[width=\linewidth]{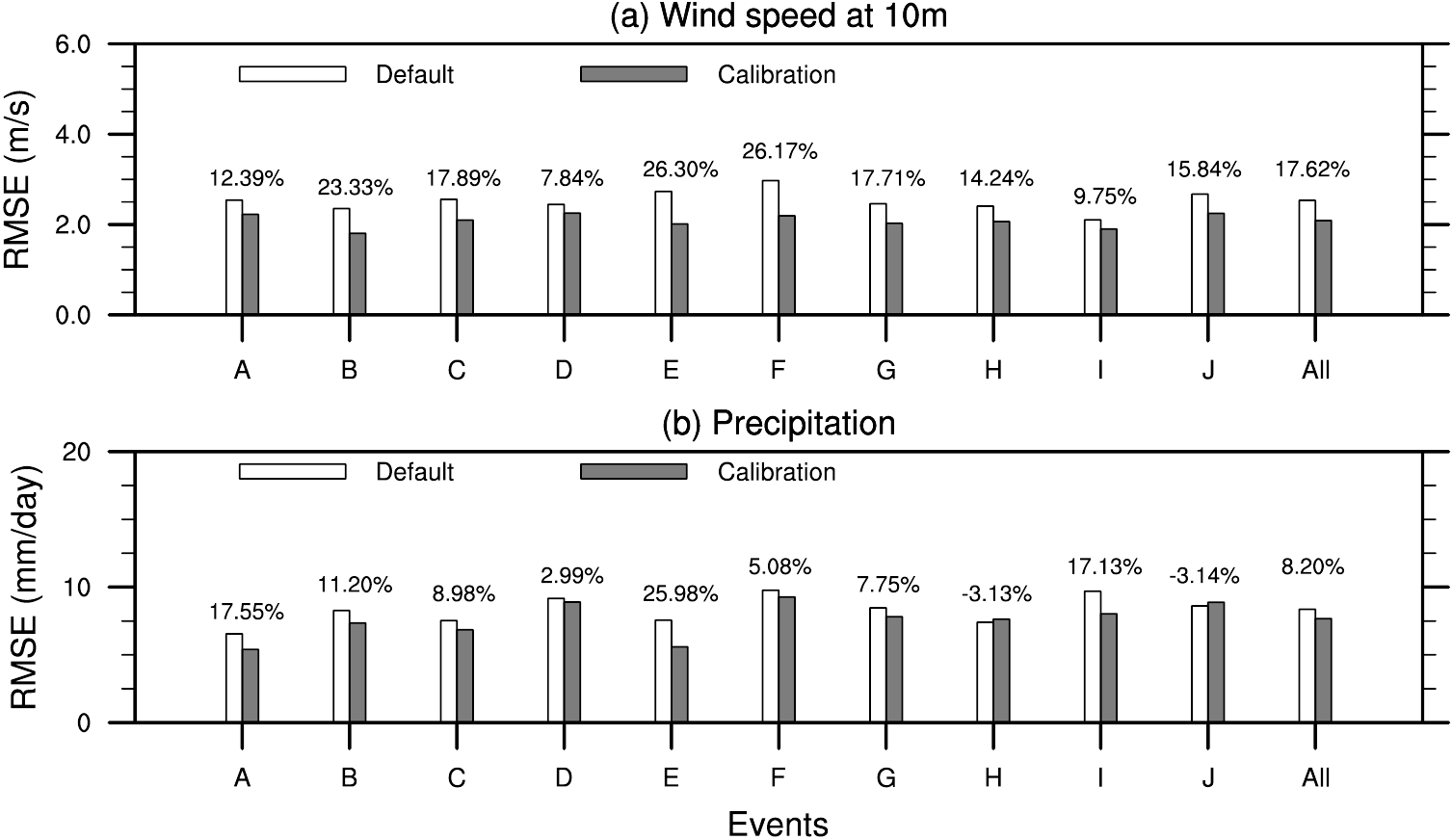}
    \caption{RMSE values of the (a) 10m wind speed (b) precipitation simulations using the calibrated parameters, in comparison with the default RMSE values, for the calibration events (A–J).}
    \label{calibration_FNL}
\end{figure}

\begin{figure}
    \centering
    \includegraphics[width=\linewidth]{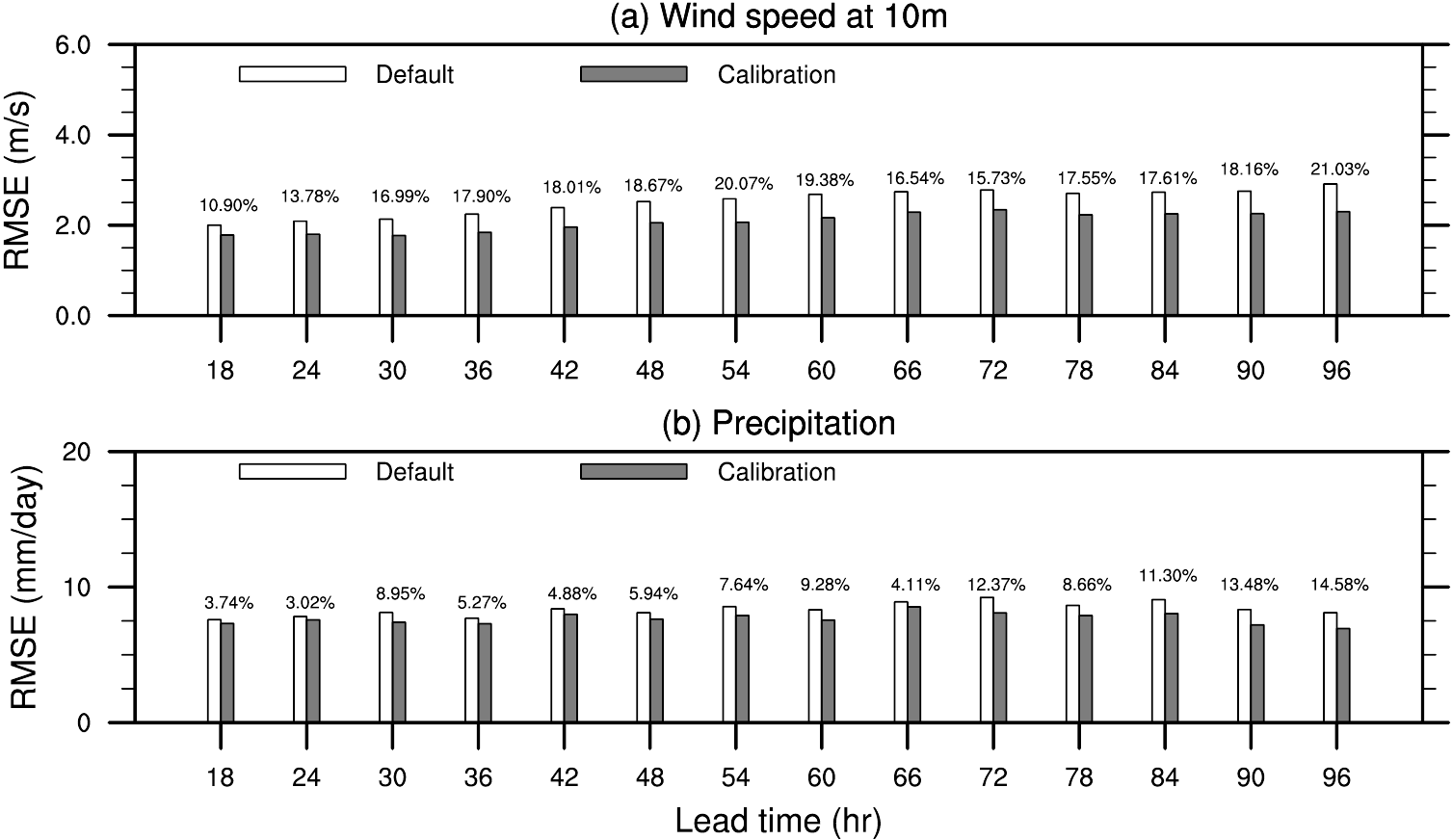}
    \caption{Comparisons of the RMSE values of (a) 10m wind speed and (b) precipitation, at 6-hourly lead times, simulated using the WRF model with default and calibrated parameters, for the calibration events (A-J).}
    \label{rmse_time_evolution_calibration}
\end{figure}

Figure \ref{calibration_FNL} shows a comparison of RMSE values of the calibrated simulations with the default simulations for the ten calibration events (A-J) over the Bay of Bengal region, which indicates a general trend of reduction in the RMSE. The reduction in the errors with the simulations of the 10m wind speed ranges from 7.84\% for event D to 26.30\% for event E, and the reduction is 17.62\% for all the events combined. Similarly, the RMSE in precipitation reduced from 2.99\% for event D to 25.98\% for event E, and the reduction is 8.20\% for all the events combined. However, some events show a marginal increase in RMSE value (-3.13\% for event H and -3.14\% for event J) for the precipitation simulation, which can be attributed to the definition of optimization objective function that focuses on the cumulative NRMSE value of all the events together rather than individual events. Thus, a few events with a small increase in the error is unavoidable. However, there is a very good overall improvement is seen. These results show that the calibrated parameters can improve the simulations of 10m wind speed and precipitation together and are consistent over the ten calibration events. In addition to the comparison of individual events, the time evolution of RMSE values for the WRF simulations with default and calibrated parameters at different lead times are also compared and shown in Figure \ref{rmse_time_evolution_calibration}. The results clearly show that the calibrated parameters improved the WRF simulations of 10m wind speed and precipitation at a 6-hourly lead time over the default simulations. The improvement in 10m wind speed simulations ranges from 10.09\% to 21.03\%, and the improvement in precipitation simulations ranges from 3.02\% to 14.58\%. 
\subsubsection{Comparison of spatial distributions for the WRF simulations with default and calibrated parameters}
Spatial distributions of the domain averaged 10m wind speed and precipitation for the ten calibration events together simulated with the default and calibration parameters are presented in supplementary Figures S1 and S2, respectively. The IMDAA dataset is used as the ground truth for 10m wind speed, and the IMERG rainfall data is used for precipitation. Figures S1(a-c) show that the 10m wind speed simulated with the default set of parameters and calibrated set of parameters have similar intensities and spatial coverage. However, the simulations from both default and calibrated parameters underestimate the intensities when compared to the observations. Figures S1d and S1e show that the default simulations have huge spatial coverage of \(\pm 1\) m/s bias over the land, 2 m/s positive bias over the BoB region, and 3 m/s positive bias over the central BoB region. In addition, the default simulations show a positive bias of more than 4 m/s over a small portion of the east coast region. In contrast, the calibrated simulations show \(\pm 1\) m/s bias over a major portion of the domain, except over the central BoB region, over which a positive bias of 2 m/s is seen. The spatial plots indicate that the calibrated parameters simulate the 10m wind speed better than the default parameters over the ocean and coastal regions, which are the most affected areas during the occurrence of a cyclone.\par
The spatial distributions of observed and simulated precipitation are presented in Figure S2(a-c). The simulations with the default and calibrated set of parameters underestimated the precipitation in terms of intensity and spatial coverage when compared with the observations. However, the calibrated parameters simulated the precipitation with better spatial coverage and intensity than the simulations from the default parameters. Figures S2d and S2e show that the significant differences between the default and calibrated simulations are seen over the Bangladesh coast, northern BoB, central BoB, and southeast BoB region. The default simulations show larger 8 mm/day positive bias structures over the Bangladesh coast, which is higher compared to the calibrated simulations. The calibrated simulation shows broad structures of 4 mm/day positive bias over the central BoB, whereas the default simulations show narrow structures of 8 mm/day positive bias. The default simulations show large 4 mm/day positive bias structures over the southeast BoB and Srilanka coast, whereas the calibrated simulations show similar structures with lesser bias. In contrast, both the simulations show similar spatial coverage of 8 mm/day negative bias over the northern BoB region. These results indicate that the calibrated parameters simulated the precipitation better than the default parameters.

\subsubsection{Comparison of Taylor statistics and SAL indices for the WRF simulations with default and calibrated parameters}
\begin{figure}
    \centering
    \includegraphics[width=0.6\linewidth]{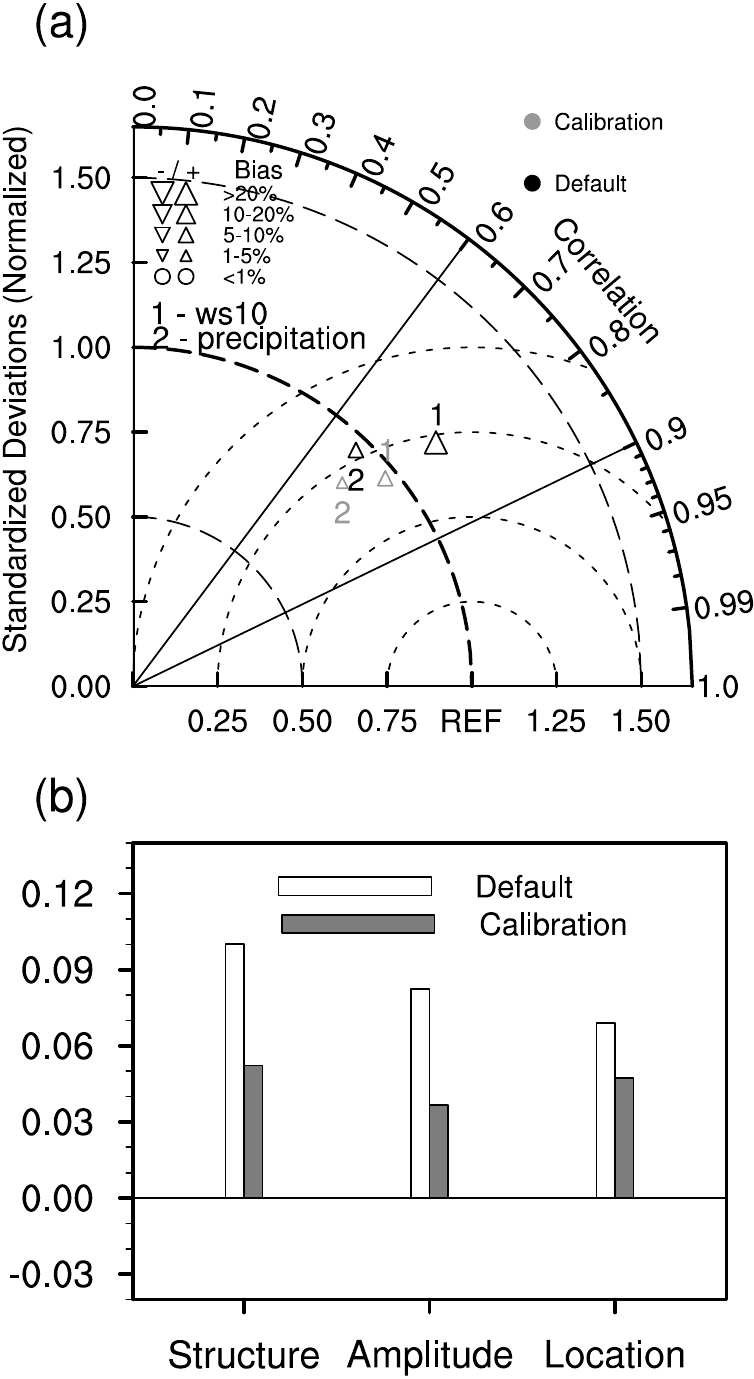}
    \caption{Comparison of (a) Taylor statistics of 10m wind speed and precipitation, (b) SAL indices of precipitation, for the simulations of calibration events (A-J) with default and calibrated parameter values.}
    \label{taylor_SAL_calibration}
\end{figure}

The spatial plots show a visual comparison that provides qualitative differences between the default and calibrated simulations. The Taylor statistics and the SAL indices are evaluated to examine the differences quantitatively and are shown in Figure \ref{taylor_SAL_calibration}. The Taylor statistics are calculated for the 10m wind speed and precipitation, whereas the SAL indices are calculated only for the precipitation as suggested by the studies of \cite{wernli2008sal}. Figure \ref{taylor_SAL_calibration}(a) show that the calibrated 10m wind speed simulations have less bias, less centered RMS error and similar variations to that of observations when compared to the default simulations. However, both the simulations have the same correlation coefficient. The calibrated precipitation simulations show a smaller bias, less centered RMS error, and a higher correlation coefficient when compared to the default simulations. The default precipitation simulations have normalized standard deviation close to the observed one, whereas the calibrated simulations have smaller variations when compared to the observations. These results show that the calibrated 10m wind speed and precipitation simulations are positioned close to the reference compared to the default simulations, which indicates that the 10m wind speed and the precipitation are better simulated with the calibrated parameter values compared to the default.\par 
The structure (S), amplitude (A), and location (L) scores of the precipitation simulated with default and calibrated parameters are presented in Figure \ref{taylor_SAL_calibration}(b). The structure value of calibrated precipitation (0.0522) is close to zero compared to the default precipitation (0.1003), which indicates that the precipitation objects of calibrated simulations are larger and have lesser variations than the observations. In contrast, the default precipitation objects are smaller in size with similar variations to that of the observations. The structure scores imply that the calibrated simulations captured the size and shape of precipitation objects better than the default simulations. The amplitude score of calibrated precipitation (0.0365) is close to zero compared to the default precipitation (0.0825), indicating that the calibrated simulations have a smaller bias than the default simulations and have captured the intensity better. Since the objective function of optimization is designed to improve the RMS error between the simulations and observations, the amplitude score reflects the optimization by improving bias and intensity. The location score of calibrated precipitation (0.0474) is also close to zero compared to the default precipitation (0.069), which indicates that the calibrated parameters simulate the precipitation with a smaller spatial deviation compared to the default parameters. These results further corroborate the superior performance of calibrated parameters over the default for the precipitation simulations. 
\subsubsection{Impact of calibrated parameters on cyclone track and intensity}
\begin{figure}
    \centering
    \includegraphics[width=\linewidth]{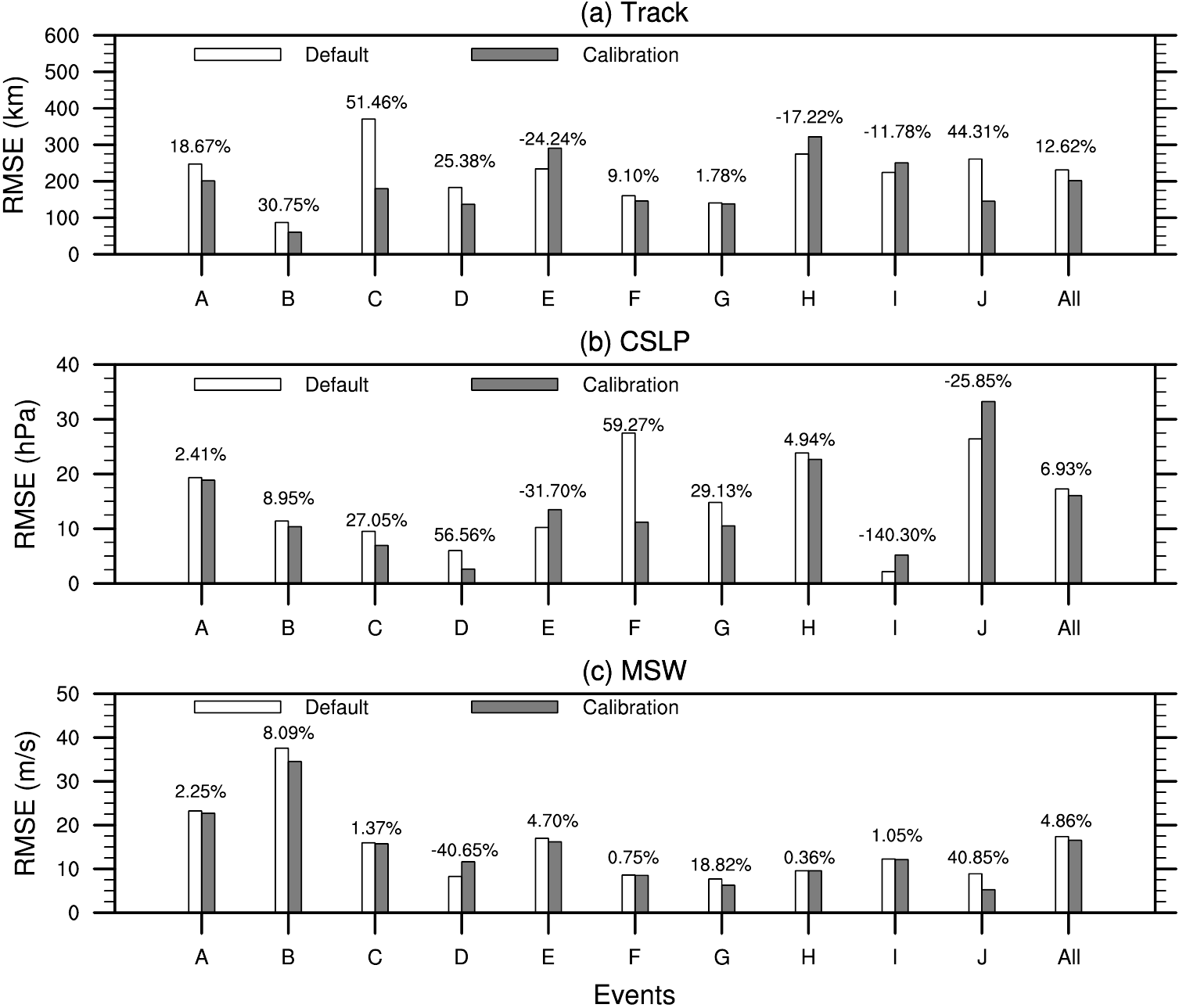}
    \caption{Comparisons of the RMSE values of (a) cyclone track, (b) central sea level pressure (CSLP), and (c) maximum sustained wind speed (MSW), at 6-hourly lead times, using the WRF model with default and calibrated parameters for the calibration events (A-J).}
    \label{track_cslp_msw}
\end{figure}

The impact of calibrated parameters on the simulations of cyclone track and intensity is examined by comparing the RMSE values between the IMD observations and the WRF simulations with default and calibrated parameters for the ten calibration events (A-J) as shown in Figure \ref{track_cslp_msw}. The relative improvement of the calibrated simulations over the default simulations is also shown over the RMSE bars. The results clearly show that the calibrated parameters improve the simulations of cyclone track, CSLP, and MSW over the default parameters with an overall improvement of 12.62\%, 6.93\%, and 4.86\%, respectively. The reduction in RMSE of cyclone track ranges from 1.78\% for event G to 51.46\% for event C. however, the simulations of calibrated parameters for the events E, H, and I show an increased RMSE, which is unavoidable since the goal here is an overall reduction in RMSE across all the events. The reduction in RMSE of CSLP ranges from 2.41\% for event A to 59.27\% for event F, and the events E, I, and J show an increased RMSE. Similarly, the reduction in RMSE of MSW ranges from 0.36\% for event H to 40.85\% for event H, while event D alone shows an increased RMSE. These results imply that the calibrated parameters better capture the cyclone track, CSLP, and MSW over the default parameters. \par
\begin{figure}
    \centering
    \includegraphics[width=\linewidth]{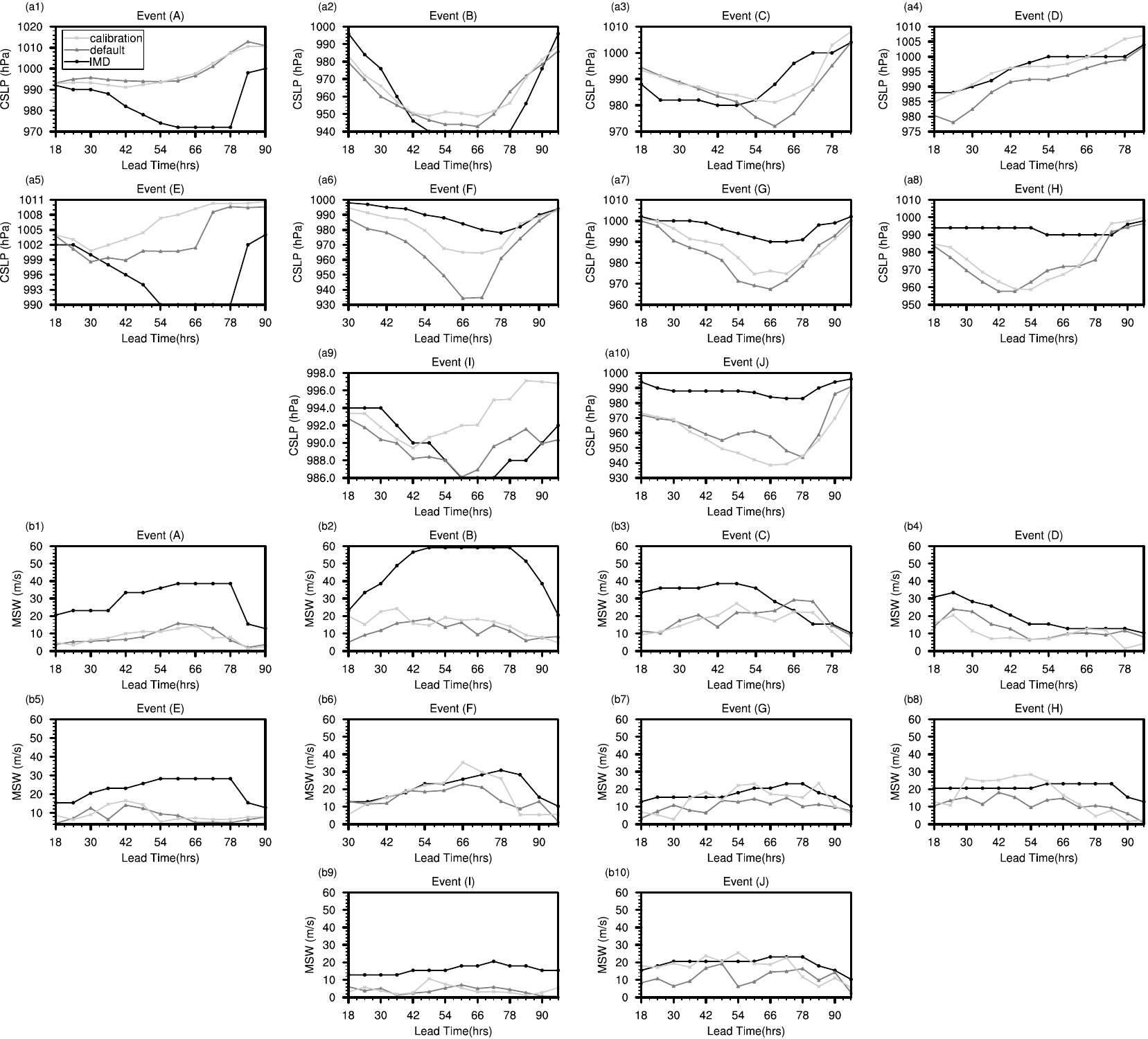}
    \caption{Comparisons of the time evolution simulations of (a1-a10) CSLP and (b1-b10) MSW, for the calibration events (A-J), using the default and calibration parameters, compared with the IMD observations.}
    \label{cslp_msw_time_evolution}
\end{figure}

A comparison of the evolution of CSLP and MSW for each event has also been conducted and shown in Figure \ref{cslp_msw_time_evolution}. The results show that the WRF simulations with default and calibrated parameters have captured the intensity evolution but have underestimated the cyclone intensity in terms of CSLP and MSW. The CSLP simulations of default and calibrated parameters show a relatively similar order of intensity for all the events except for F, for which the default parameters overestimated the intensity with a deviation of 50 hPa, whereas the calibrated parameters overestimated the intensity with a deviation of 10 hPa. For the MSW simulations, the calibrated parameters show a little better performance than the default parameters for every event, except for D, for which the calibrated simulations show a deviation of 20 m/s, whereas the default simulations show a deviation of 10 m/s. These results suggest that the overall performance of calibrated parameters is relatively better than the default parameters.
\subsubsection{Impact of calibrated parameters on the atmospheric structure}
\begin{figure}
    \centering
    \includegraphics[width=\linewidth]{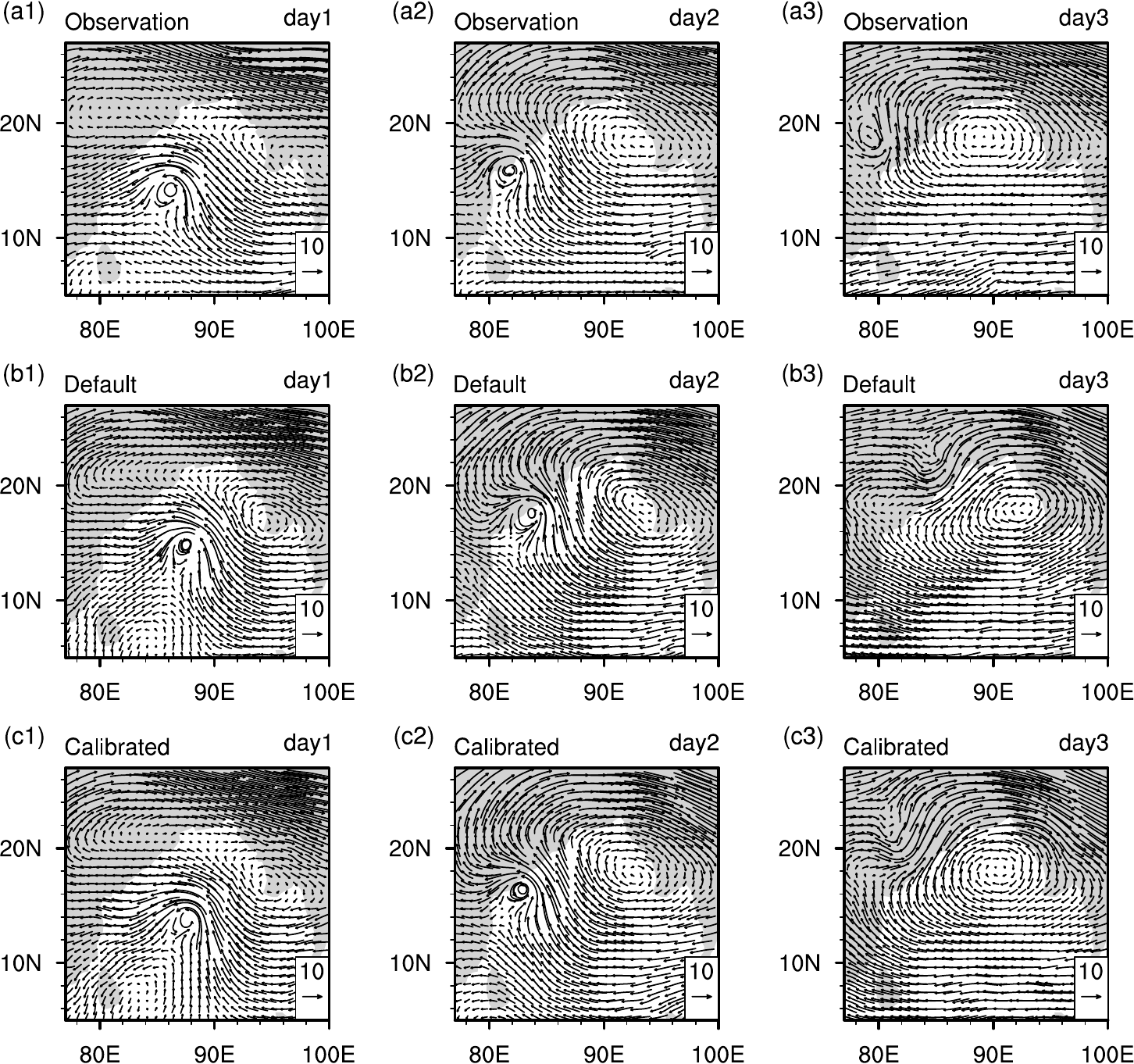}
    \caption{An illustration of the wind velocity field at 500 hPa for the simulation of VSSC Leher. (a1-a3) show observations at the end of day1, day2, and day3; (b1-b3) show the simulations with default parameters; and (c1-c3) show the simulations with calibrated parameters.}
    \label{velocity_field_event3}
\end{figure}

\begin{figure}
    \centering
    \includegraphics[width=\linewidth]{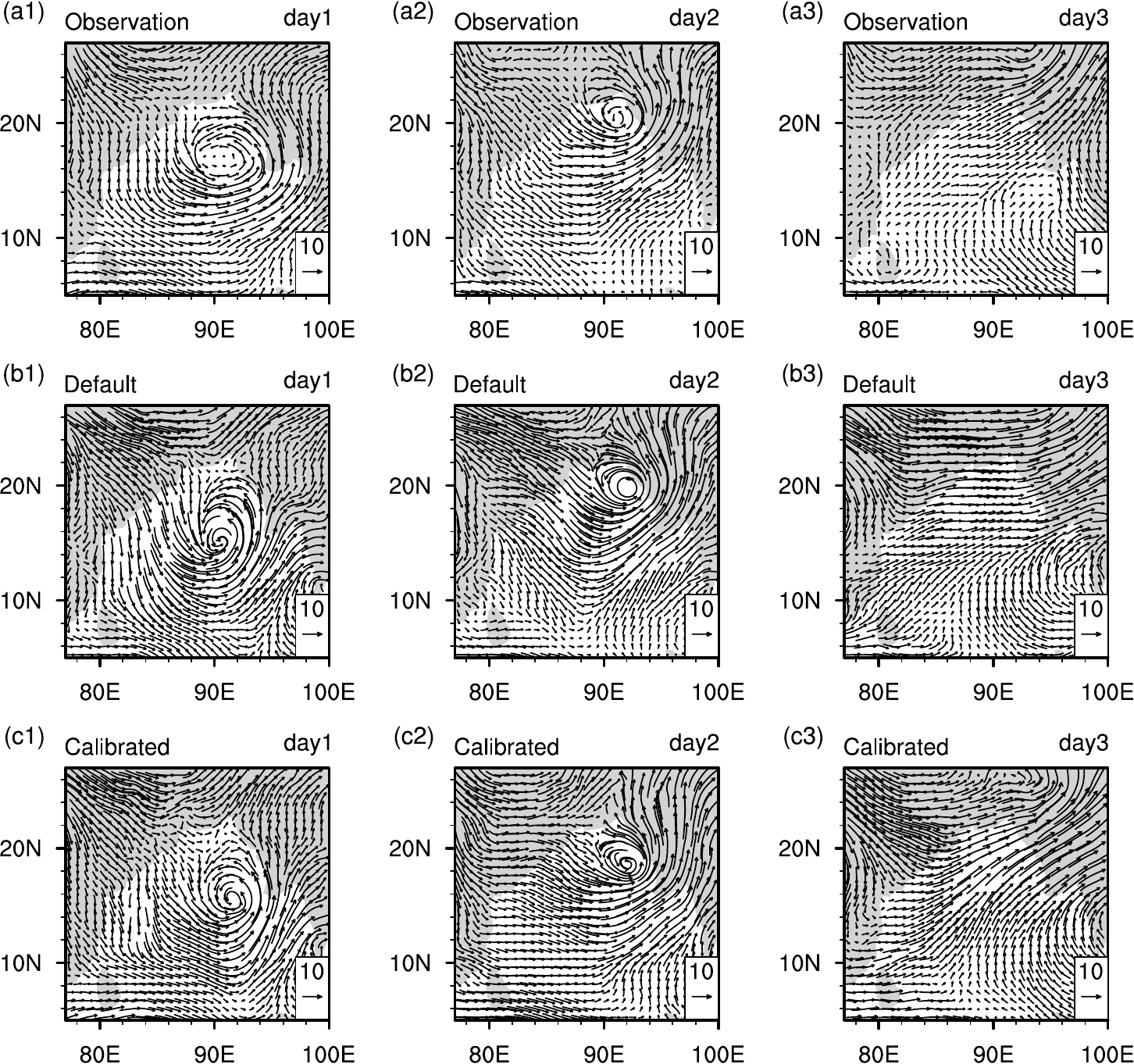}
    \caption{An illustration of the wind velocity field at 500 hPa for the simulation of SCS Mora. (a1-a3) show observations at the end of day1, day2, and day3; (b1-b3) show the simulations with default parameters; and (c1-c3) show the simulations with calibrated parameters.}
    \label{velocity_field_event6}
\end{figure}

The objective function of the calibration methodology is designed to minimize the deviation between observations and simulations of wind speed and precipitation at the surface level. However, the calibration of surface-level fields will have an impact on the upper atmospheric variables positively or negatively. The impact of calibration on the higher-level atmospheric structure is therefore examined by observing the velocity field at 500 hPa for event 3 (cyclone Leher) and event 6 (cyclone Mora). Figure \ref{velocity_field_event3} show the 500 hPa velocity field of cyclone Leher at the end of \(1^{st}, 2^{nd}\) and \(3^{rd}\) days, simulated by the default and calibrated parameters and compared with the observations. The observed velocity fields clearly show a cyclonic circulation at the end of all days and an anticyclonic circulation at the end of \(2^{nd}\) and \(3^{rd}\) days. The default parameters simulated a cyclonic circulation at the end of \(1^{st}\) and \(2^{nd}\) days with an overestimated intensity compared to the observations and an anticyclonic circulation at the end of all days. However, the locations of cyclonic and anticyclonic circulations have deviated from the observations, and the structure of the anticyclonic circulation is also different from the observations. The calibrated parameters simulated a cyclonic circulation at the end of all days with a relatively similar intensity to observations and an anticyclonic circulation at the end of all days with a similar structure to that of observations. The cyclonic and anticyclonic circulations are also very close to the observations compared to the default simulations. Similarly, Figure \ref{velocity_field_event6} shows a comparison of 500 hPa velocity field between the observations and simulations with default and calibrated parameters, for cyclone Mora, at the end of \(1^{st}, 2^{nd}\) and \(3^{rd}\) days. The observations clearly show a cyclonic circulation at the end of \(1^{st}, 2^{nd}\) and \(3^{rd}\) days. The default parameters simulated a cyclonic circulation at the end of \(1^{st}\) and \( 2^{nd}\) days with an overestimation compared to the observations. However, there is no circulation simulated at the end of \(3^{rd}\) day. The calibrated parameters simulated a cyclonic circulation at the end of all days with lesser intensity than the default parameters. Although the locations of the cyclonic circulation at the end of \(1^{st}\) and \(2^{nd}\) simulated by the default and calibrated parameters are very close to the observations, the intensity of the circulation simulated by the calibrated parameters simulated is better than the default parameters. These results indicate that the calibration indeed improves the 500 hPa level atmospheric structure. 

\subsection{Verifying the robustness of the calibrated parameters for other events}
To corroborate the robustness of the calibrated parameters, eight new tropical cyclone events, which are different from the calibration events and are referred to as the validation events hereafter are simulated using the WRF model. The details of the validation events are presented in Table \ref{cyclones} and the corresponding IMD observed tracks are shown in Figure \ref{IMD_tracks}(b). The validation events are also simulated using the same configuration as the calibration experiments, including domain configuration, simulation duration, physics schemes, and boundary data. \par
\begin{figure}
    \centering
    \includegraphics[width=\linewidth]{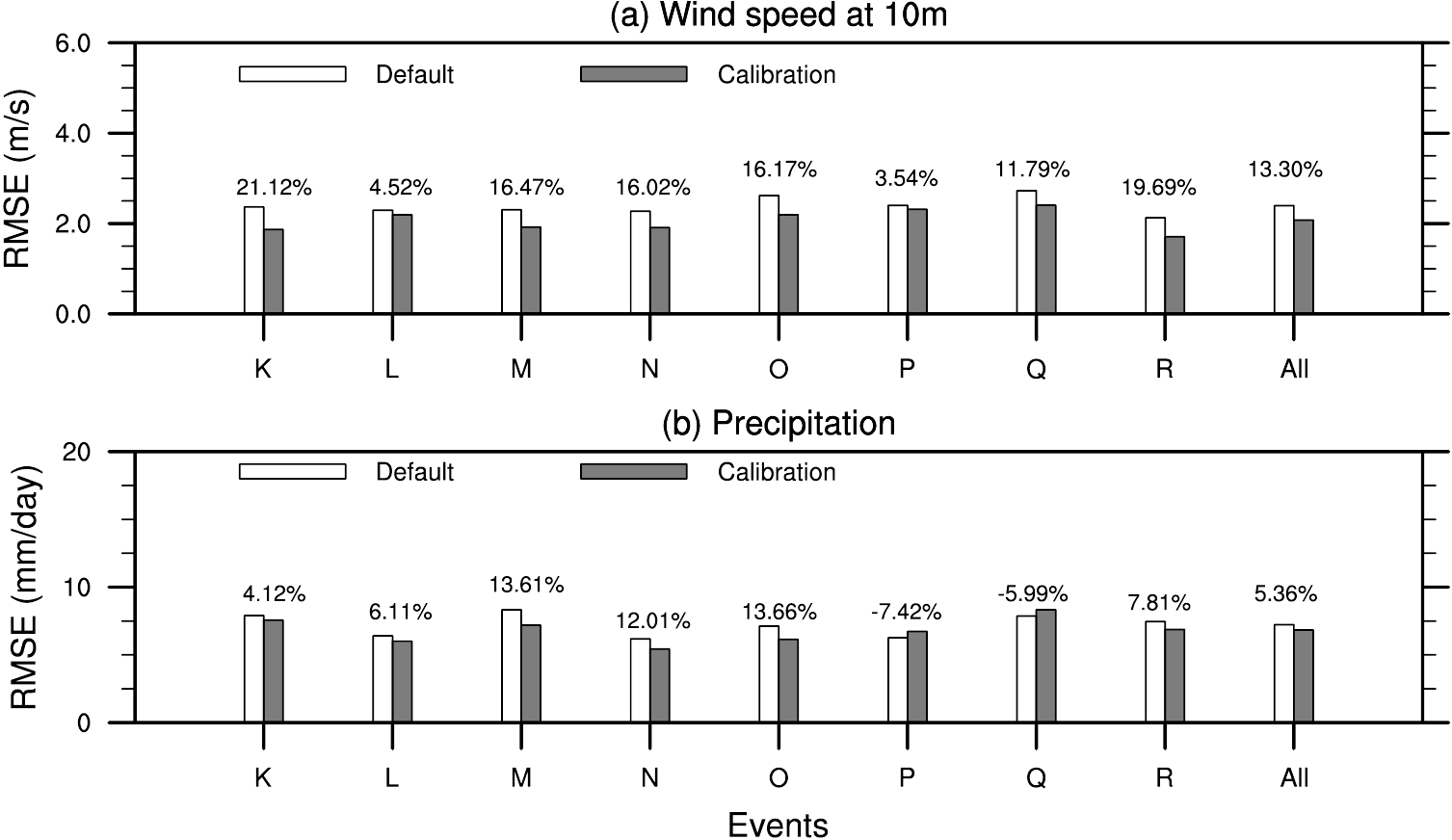}
    \caption{Comparison of the RMSE values of (a) 10m wind speed and (b) precipitation, simulated using the WRF model with default and calibrated parameters, for the validation events (K-R).}
    \label{validation_FNL}
\end{figure}

Figure \ref{validation_FNL} shows a comparison of RMSE values of the WRF simulations with the calibrated and the default parameters for the eight validation events (K-R) over the Bay of Bengal region, which indicate a general trend of reduction in the RMSE. The reduction in the error of 10m wind speed in terms of the RMSE ranges from 3.54\% for event P to 21.12\% for event K, and the reduction in the overall RMSE is 13.30\%. Similarly, the RMSE reduction for precipitation ranges from 4.12\% for event K to 13.66\% for event O, and the overall reduction is 5.36\% for all the events considered together. It is apparent that the calibrated parameters improved the simulations of calibration events as well as the validation events. Similar to the calibration events, the time evolution of wind speed and precipitation (figures not shown) simulated by the calibrated parameters show an improving trend over the default parameters for the validation events.\par
\begin{figure}
    \centering
    \includegraphics[width=0.6\linewidth]{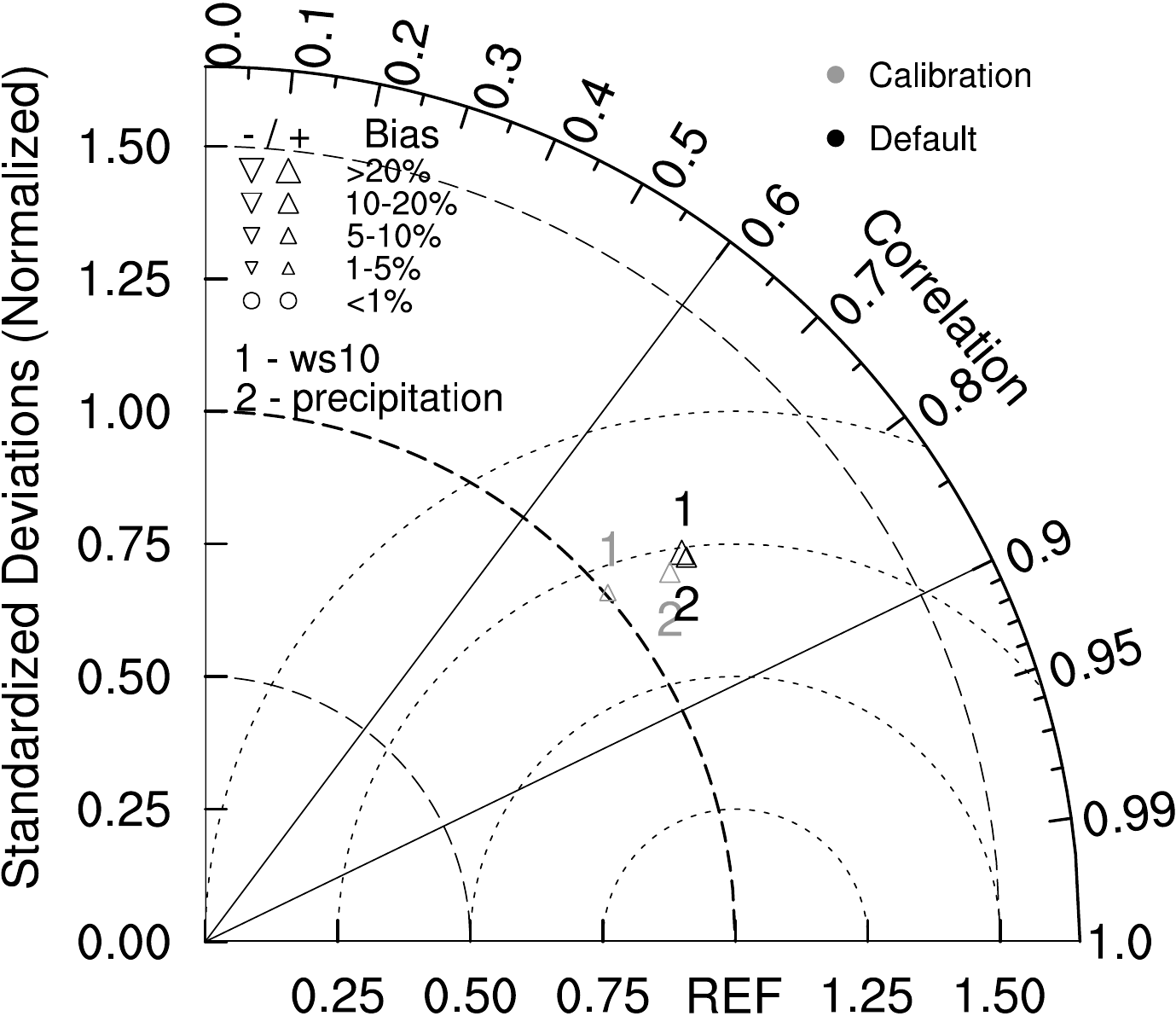}
    \caption{Comparison of Taylor statistics of 10m wind speed and precipitation for the simulations of validation events (K-R) using the default and calibrated parameter values.}
    \label{taylor_validation}
\end{figure}

Spatial plots of the domain averaged 10m wind speed and precipitation simulations (figures not shown) indicate that the calibrated parameters simulated the variables close to observations with less bias than the default simulations. The Taylor statistics are evaluated for the validation events to examine further the quantitative differences between the default and calibrated simulations, and are presented in Figure \ref{taylor_validation}. The Taylor statistics of 10m wind speed show that the calibrated simulations have lesser bias and RMS errors than the default simulations and a standard deviation similar to that of the observed one. In contrast, the default 10m wind speed simulations have a lesser correlation coefficient than the calibrated simulations and a standard deviation higher than that of the observed one. Similarly, the Taylor statistics of precipitation show that the calibrated simulations have lesser RMS error and higher correlation coefficient than the default simulations, a standard deviation close to that of the observed one, and a bias similar to that of default simulations. The Taylor statistics of 10m wind speed and precipitation simulated by the calibrated parameters are positioned close to the reference than the default simulations, indicating the better performance of the calibrated parameters.\par
\subsection{Impact of calibrated parameters on simulations with different boundary conditions}
To examine the robustness of the calibrated parameters across different boundary conditions, the ECMWF Re-Analysis (ERA5) data with a \(1.0^{\circ}\) grid is used to drive the WRF simulations of the calibration and validation events (A-R). The simulation experiments adopted the configurations that of the calibration experiments, such as model configuration, simulation area, and simulation duration, except for the boundary conditions, for which the ERA5 data are used. The performance of the calibrated parameters is evaluated by comparing the RMSE values of 10m wind speed and precipitation.\par
\begin{figure}
    \centering
    \includegraphics[width=\linewidth]{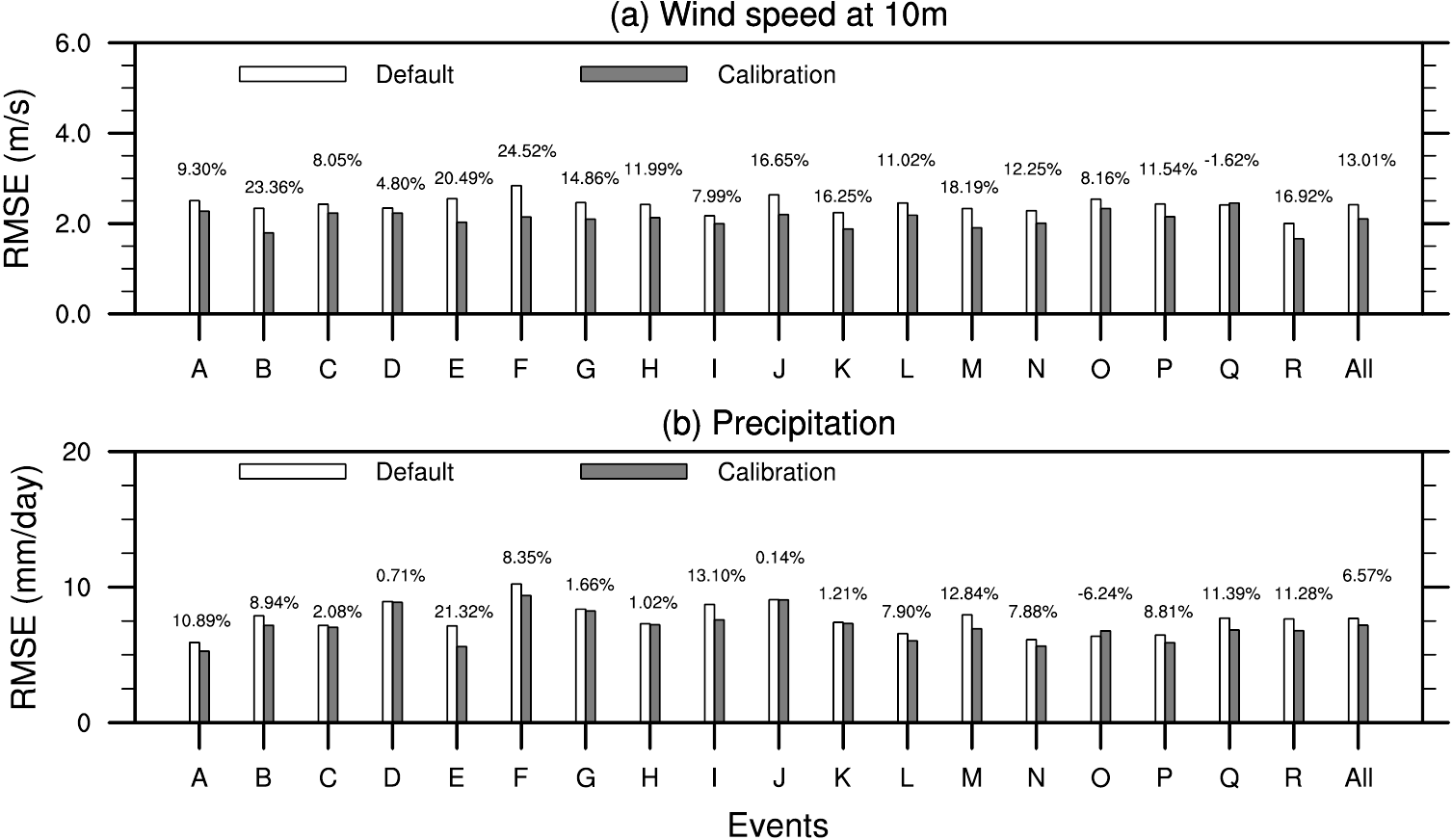}
    \caption{Comparison of RMSE values of (a) 10m wind speed and (b) precipitation, simulated using the default and calibration parameters for the observed events (A-R), using the 1deg ERA5 reanalysis data as boundary conditions with nested domain at 12km inner domain resolution.}
    \label{ERA5_1deg}
\end{figure}

 
Figure \ref{ERA5_1deg} presents the RMSE values of 10m wind speed and precipitation simulations with default and calibration parameters, for the total events (A-R), by considering the ERA5 data as the boundary conditions. The results clearly show a better performance of the calibrated parameter values over the default ones by reducing the RMSE values. The reduction in the RMSE of the 10m wind speed ranges from 4.80\% for event D to 24.52\% for event F, and the overall reduction is 13.01\% for all the events considered together. Similarly, the RMSE reduction for precipitation ranges from 0.14\% for event J to 21.32\% for event E, and the overall reduction is 6.57\% for all the events considered together. Although the events Q and O show an increase in RMSE of 1.62\% and 6.24\% for the 10m wind speed and precipitation simulations, respectively, the overall reduction compensated for this increment. From these results, it can be concluded that the calibrated parameter values still simulate better forecasts of 10m wind speed and precipitation compared to the default parameter values for the simulations of tropical cyclones over the BoB region even if the boundary conditions are replaced from FNL to ERA5. 
\subsection{Impact of calibrated parameters on simulations with different grid resolutions}
To investigate the robustness of the calibrated parameters for the simulations of tropical cyclones over the BoB region at different spatial resolutions, two sets of WRF simulations with grid resolutions different from the calibration experiments are conducted to simulate the observed events (A-R). The experiments include single domain simulations with a 12 km grid resolution and nested domain simulations with a 12 km outer domain resolution and a 4 km inner domain grid resolution. The number of grid points in the outer domain (d01) is \(520 \times 520\), and the inner domain (d02) is \(960 \times 960\). The WRF model domain configuration of these two experiments is illustrated in supplementary Figure S3. These simulations differ from the calibration simulations. The change in grid resolution leads to a change in the number of grid points to maintain the simulation area similar to that of the calibration experiments. Since the parent domain grid resolution is at 12 km, the driving boundary conditions taken at \(0.25^{\circ}\) grid is a best practice. However, there is no FNL data at \(0.25^{\circ}\) is available before the 2014 period, which is required for the simulations of the selected tropical cyclones. Thus, the ERA5 data with a \(0.25^{\circ}\) grid is used to drive the WRF simulations for both experiments. The remaining configurations, including physics schemes and simulation duration, are similar to that of the calibration and validation simulations. The performance of the calibrated parameters is evaluated by comparing the RMSE values of 10m wind speed and precipitation.\par 

\begin{figure}
    \centering
    \includegraphics[width=\linewidth]{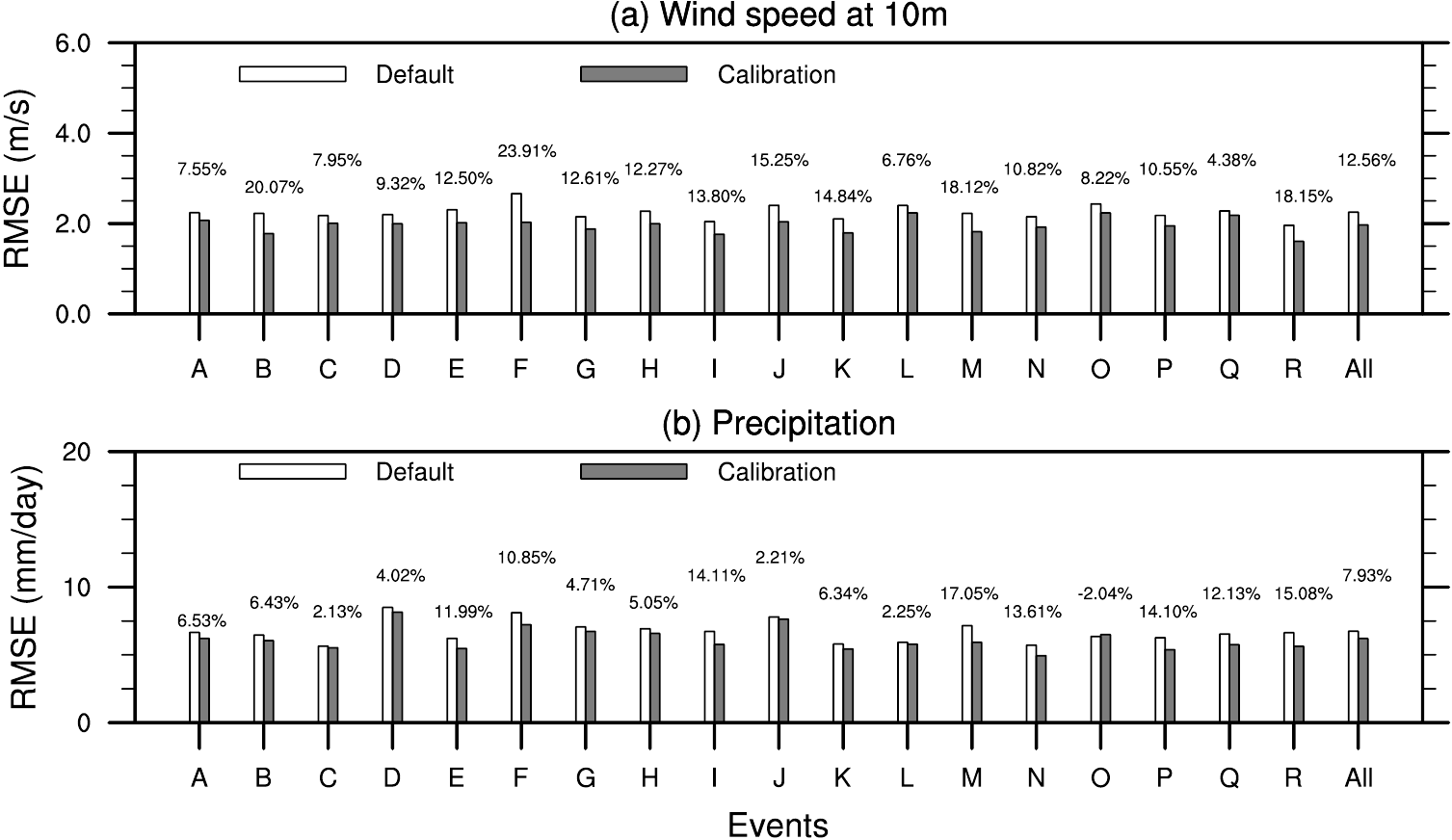}
    \caption{Comparison of RMSE values of (a) 10m wind speed and (b) precipitation, simulated using the default and calibration parameters for the observed events (A-R), using the 0.25 deg ERA5 reanalysis data as boundary conditions with a single domain at 12 km resolution.}
    \label{ERA5_0.25deg_1dom}
\end{figure}

Figure \ref{ERA5_0.25deg_1dom} presents a comparison of RMSE values of 10m wind speed and precipitation simulations with default and calibration parameter values on a 12 km single domain for the observed simulation events (A-R). The calibration parameters have better 10m wind speed and precipitation simulation capabilities than those with the default parameters. The reduction in the RMSE of the 10m wind speed ranges from 4.38\% for event Q to 23.91\% for event F, and the overall reduction is 12.56\% for all the events considered together. Similarly, the RMSE for precipitation reduces from 2.13\% for event C to 17.05\% for event M, and the overall reduction is 7.93\% for all the events considered together. The 10m wind speed simulations show that calibrated parameters decreased the RMSE values for all the events. However, the overall improvement is a little less than the calibration and validation experiments together (15.79\%). Similarly, the precipitation simulations show that the calibrated parameters decreased the RMSE values for all the events except event O, and the overall improvement is even higher than the calibration and validation experiments together (which is 7.13\%). These results suggest that calibrated parameters can improve the 10m wind speed and precipitation simulations on a 12 km single domain with ERA5 boundary data at \(0.25^{\circ}\) grid and demonstrate the superior performance of the calibration parameters over the default parameters. \par
\begin{figure}
    \centering
    \includegraphics[width=\linewidth]{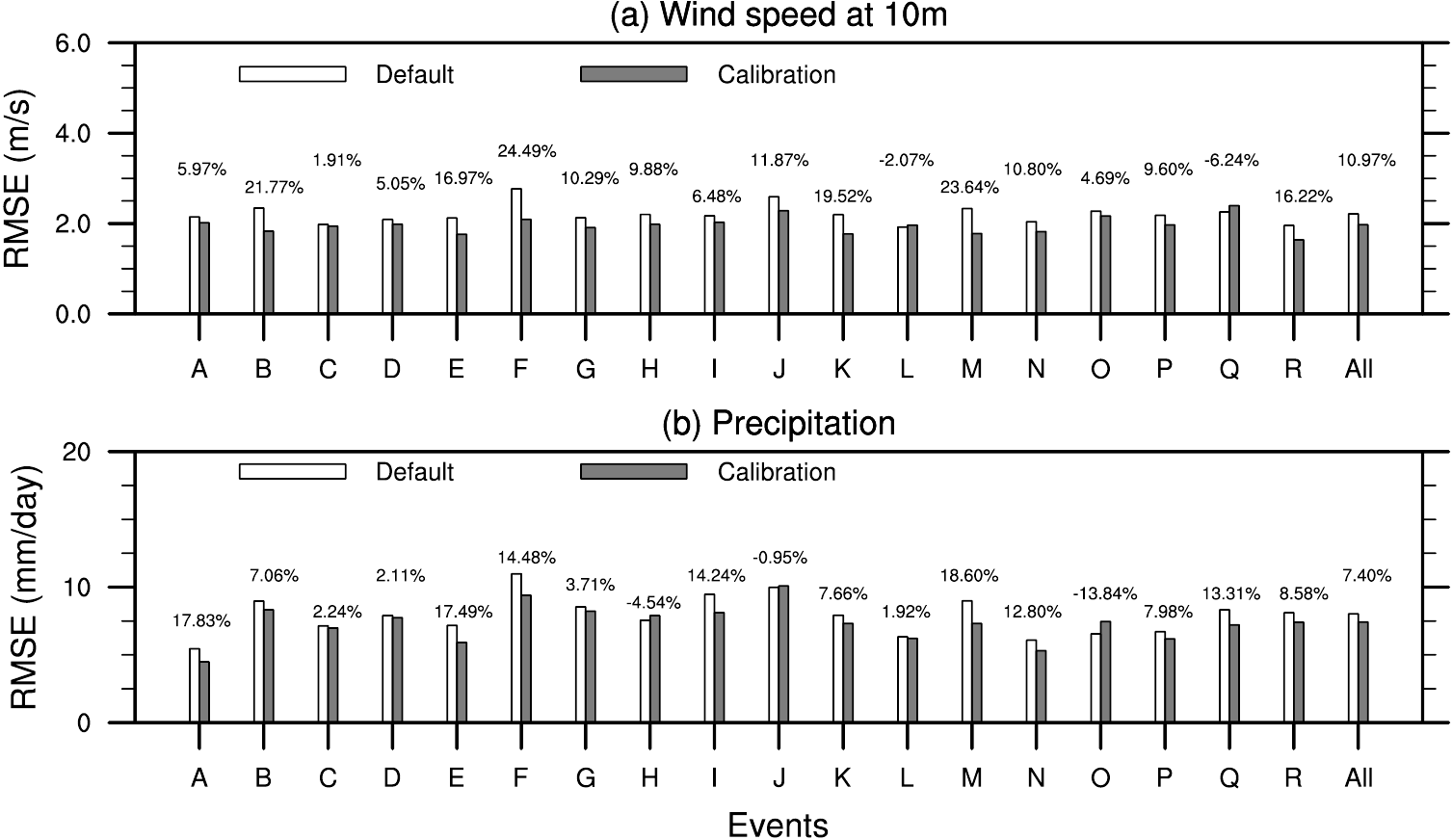}
    \caption{Comparison of RMSE values of (a) 10m wind speed and (b) precipitation, simulated using the default and calibration parameters for the observed events (A-R), using the 0.25 deg ERA5 reanalysis data as boundary conditions with nested domain at 4 km inner domain resolution.}
    \label{ERA5_0.25deg_2dom}
\end{figure}

For the 4 km grid resolution nested domain experiments, Figure \ref{ERA5_0.25deg_2dom} presents a comparison of RMSE values of 10m wind speed and precipitation simulations with default and calibration parameter values for the observed simulation events (A-R). The reduction in the RMSE of the 10m wind speed ranges from 1.91\% for event C to 24.49\% for event F, and the overall reduction is 10.97\% for all the events considered together. Similarly, the RMSE for precipitation reduces from 1.92\% for event L to 18.60\% for event M, and the overall reduction is 7.40\% for all the events considered together. The 10m wind speed simulations show that calibrated parameters decreased the RMSE values for all the events except events L and Q, and the overall improvement is far less compared to the calibration and validation experiments together (which is 15.79\%). Similarly, the precipitation simulations show that the calibrated parameters decreased the RMSE values for all the events except events H, J, and O, and the overall improvement is a little higher than the calibration and validation experiments together. Though the performance of the calibrated parameters is better than the default parameters, the improvement in RMSE is lesser compared to that of calibration and validation experiments. As a whole, the performance of the calibrated parameters is superior to that of the default parameters for the simulations of tropical cyclones over the BoB region, even at 12km single domain simulations and 4km nested domain simulations. 
\subsection{Physical interpretation of the calibrated parameter values}
The parameter values obtained from the described optimizations experiments and the default values are listed in Table \ref{final_parameter_values}, which indicates that the parameters have highly inconsistent variations among them. Compared with default values, the calibrated parameter values are increased for znt\_zf, ice\_stokes\_fac, and pfac, whereas the values decreased for pe, cssca\_fac, secang, and porsl, and the value of Karman is unchanged. It is impractical to constitute a straightforward relationship between the parameters and the simulated model output variables because of the nonlinear interactions present among the various physical processes \citep{di2018assessing,chinta2020calibration}. \par
\begin{table}
    \caption{Final parameter values obtained from the optimization of 10m wind speed, precipitation, and the calibrated parameters.}
    \scriptsize
    \centering
    \renewcommand\arraystretch{1.75}
    \begin{tabular}{p{0.15\linewidth}p{0.15\linewidth}p{0.15\linewidth}p{0.15\linewidth}p{0.15\linewidth}}
    \hline
    \textbf{Parameter} & \textbf{Default}   & \textbf{Opt\_ws10} & \textbf{Opt\_rain} & \textbf{Calibration} \\
    \hline
    \hline
    znt\_zf	&	1	&	2	&	0.5293	&	2	\\
    karman	&	0.4	&	0.42	&	0.3503	&	0.4	\\
    pe	&	1	&	0.5001	&	0.5123	&	0.5001	\\
    ice\_stokes\_fac	&	14900	&	29999.7672	&	28500.5724	&	29915.6267	\\
    cssca (cssca\_fac)	&	1.00E-05	&	5.00E-06	&	1.99E-05	&	5.00E-06	\\
    Secang	&	1.66	&	1.7474	&	0.5019	&	0.5041	\\
    porsl	&	1	&	0.5	&	1.8594	&	0.5011	\\
    pfac	&	2	&	3	&	2.9893	&	2.9999	\\
    \hline
    \end{tabular}
    \label{final_parameter_values}
\end{table}
The znt\_zf parameter (scaling related to surface roughness) is used in the surface layer scheme, which directly controls the roughness length and affects the surface wind speed. A higher znt\_zf value means a higher roughness length which results in decreased 10m wind speed. The Karman parameter (Von Karman constant) is used in both surface layer and planetary boundary layer schemes, which governs the flow speed profile in a wall-normal shear flow and influences the bulk transfer coefficient of momentum, heat, moisture, and diffusivity coefficient of momentum. A higher Karman value can increase the exchange coefficient of momentum (surface drag coefficient) and decrease the surface wind speed. At the same time, a higher Karman value can increase the bulk transfer coefficient of moisture that leads to an increase in precipitation. The parameter Pe (multiplier for entrainment mass flux rate) is used in the cumulus physics scheme, which regulates the entraining of ambient air into the updraft air parcel. A lower Pe value indicates a smaller amount of entrainment, which makes the atmosphere more unstable and helps in the formation of deep convective clouds, which ultimately leads to heavy convective precipitation. The ice\_stokes\_fac (scaling factor applied to icefall velocity) is used in the microphysics scheme, which controls the terminal fall velocity of cloud ice particles. A higher ice\_stokes\_fac value increases the sedimentation of ice crystals and increases the conversion from cloud ice to rainwater, resulting in increased precipitation. The cssca\_fac is used in the shortwave radiation scheme, which governs the attenuation of downward shortwave radiation. A smaller value of cssca\_fac leads to less attenuation and scattering of solar radiation, resulting in a higher amount of solar radiation reaching the Earth's surface, which increases the evaporation at the surface that causes an increase in precipitation. The parameter secang (diffusivity angle for cloud optical depth computation) is used in the longwave radiation scheme, which governs the attenuation of the longwave radiation irradiated by the clouds towards the Earth's surface. Similar to the csscs\_fac, a smaller value of secang results in less attenuation, increasing the amount of longwave radiation reaching the Earth's surface, which ultimately results in higher precipitation. The parameter porsl (multiplier for the saturated soil water content) is used in the land surface scheme, which plays a vital role in heat exchange between land and atmosphere through moisture evapotranspiration in soil. A high value of porsl increases the evaporation of soil water, leading to an increase in precipitation. The parameter pfac (profile shape exponent for calculating the momentum diffusivity coefficient) is used in the planetary boundary layer scheme, contributing to the diffusion of turbulence eddies in the boundary layer. A higher value of pfac increases the vertical momentum diffusion and thereby increases the wind speed. In addition, a higher value of pfac also increases the heat diffusion, which makes more evaporation of water vapor from the ground and leads to higher precipitation. Because all of the parameters have a very nonlinear relationship with the model output variables, explaining the calibrated parameter values becomes quite difficult. Nevertheless, the calibrated parameters obtained in this study improved the prediction of precipitation and wind speed for the simulations of eighteen tropical cyclones over the BoB region compared to the default parameter values. 

\section{Conclusions}
The present study employed the MultiObjective Adoptive Surrogate Model-based Optimization (MO-ASMO) framework to calibrate the eight sensitive parameters of the WRF model to reduce the simulations simulation errors of 10m wind speed and precipitation for the tropical cyclones over the Bay of Bengal (BoB) region. The parameters were sampled using the Quasi Monte-Carlo (QMC) Sobol' design to create the initial parameter sets, using which the WRF model simulations of the calibration events were performed. Two surrogate models of Gaussian Process Regression (GPR) were constructed, each one for 10m wind speed and precipitation. The surrogate models were optimized using the Genetic Algorithm (GA) in an iterative process. The optimized parameter values of 10m wind speed improved the 10m wind speed simulations by 20.56\%, and that of the precipitation improved the precipitation simulations by 11.30\%. After this, the NSGA-II algorithm was employed as a multiobjective optimization and a Pareto front containing a set of non-dominated solutions of optimal parameters values was obtained. The final best solution is sorted by utilizing the TOPSIS method. The described calibration methodology yields three sets of optimal parameter values: one set minimizes the NRMSE values of 10m wind speed, another set minimizes the NRMSE values of precipitation, and the calibration set minimizes the NRMSE values of 10m wind speed and precipitation together. \par
The performance of the calibrated parameter values over the default values was evaluated by comparing the NRMSE values, spatial distributions, Taylor statistics, and the SAL indices for the simulations of calibration events. The results showed that the calibrated parameters improved the 10m wind speed and precipitation simulations by 17.62\% and 8.20\%, respectively, and outperformed the default parameters in every other aspect. A comparison of the cyclone track and intensity simulations showed that the calibrated parameters reduced the track error by 12.62\%, CSLP error by 6.93\%, and MSW error by 4.86\%. In addition to the surface fields, an examination of the wind velocity field at 500 hPa revealed that the simulations carried out with the calibrated parameters were close to the observations. Further, the performance of the calibrated parameter values was validated for a different set of cyclone events and found that an improvement of 13.30\% and 5.36\% observed for 10m wind speed and precipitation. Finally, the robustness of the calibrated parameter values across different boundary conditions and grid resolutions was also examined by taking the boundary conditions from 1deg and 0.25deg ERA5 reanalysis data, and conducting the WRF simulations at 12km single domain and 4km nested domain resolutions. The calibrated parameters obtained in this study were found to be robust across different precipitation events, driving data, and domain resolutions.

\clearpage
\acknowledgments
\noindent The model simulations are performed on Aqua High-Performance Computing (HPC) system at the Indian Institute of Technology Madras (IITM), Chennai, India, and the Aaditya HPC system at the Indian Institute of Tropical Meteorology (IITM), Pune, India. The authors would like to thank the “Centre of Excellence (CoE) in climate change impact on coastal infrastructure and the adaptation strategies” for their suggestions. Department of Science and Technology is funding the research grant DST/CCP/CoE/141/2018(C) for this project (CIE1819265DSTXSACI) under SPLICE—Climate Change Programme. The authors declare that there are no conflict of interest to disclose.

%
%
\datastatement
\noindent The WRF model main source code is modified according to the author's requirement to enable the parameter values given as inputs through the namelist.input file. The modified source code and the namelist.input files that are used for the simulations can be downloaded from the repository at \url{https://doi.org/10.5281/zenodo.5150524}\citep{baki_harish_2021_5150524}. The WRF model is built in FORTRAN 90 language. The software can be installed in any Linux based system. The PYTHON language based Application Programming Interfaces (API) are used for the machine learning algorithms namely, Gaussian Process Regression (GPR) (\url{https://scikit-learn.org/stable/modules/generated/sklearn.gaussian_process.GaussianProcessRegressor.html#}), Genetic Algorithm (GA) (\url{https://github.com/rmsolgi/geneticalgorithm}), and Non-dominated Sorted Genetic Algorithm-II (NSGA-II) (\url{https://pymoo.org/}). These APIs are publicly available and the contact information can be found at the respective websites.

\bibliographystyle{ametsocV6}
\bibliography{references}

\end{document}





\begin{figure}
    \centering
    \includegraphics[width=0.8\linewidth]{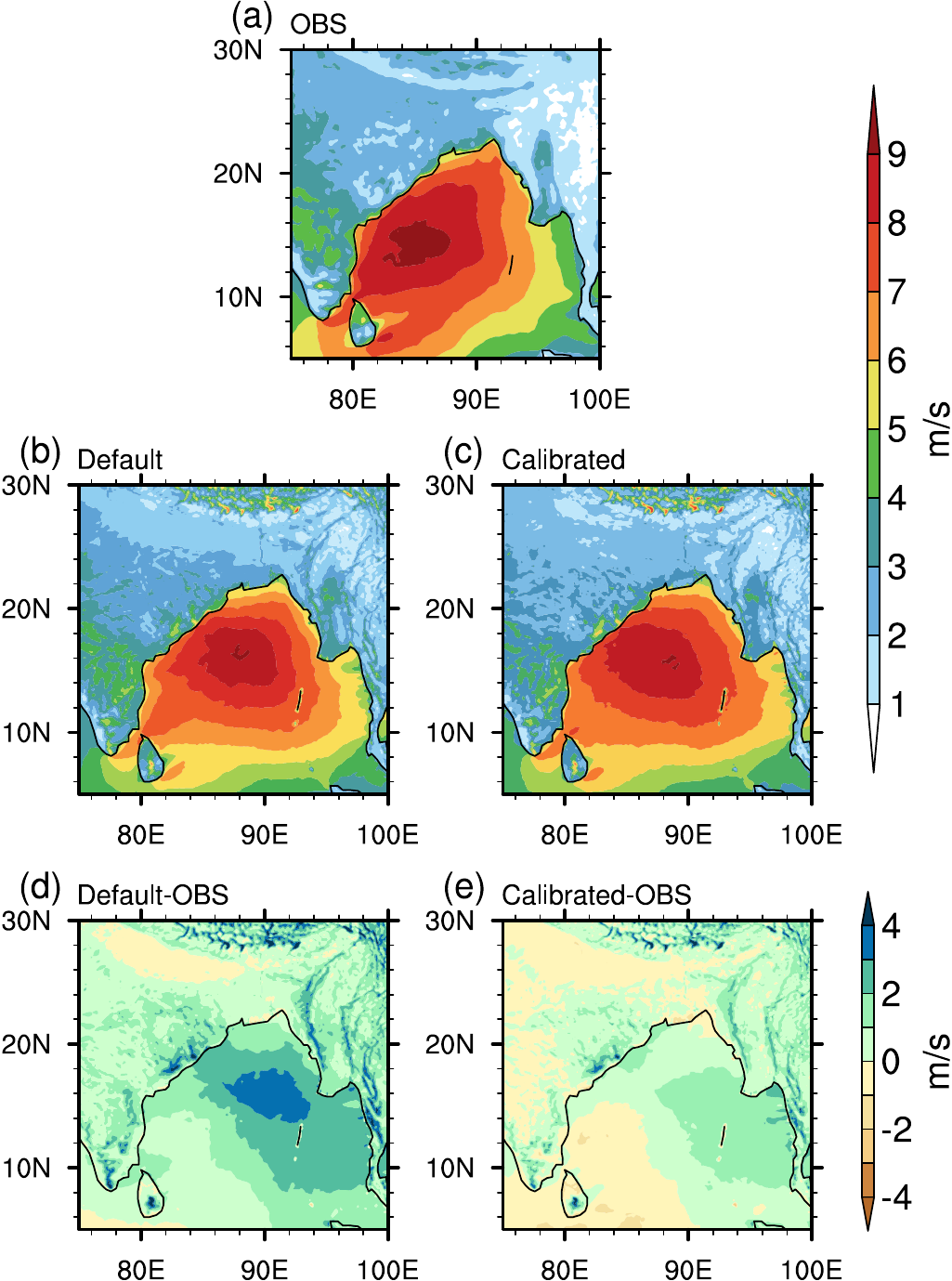}
    \caption{Spatial distributions of the domain averaged 10m wind speed simulated using the default and calibrated parameters, in comparison with the observations, for the calibration events (A-J). (a) observed wind speed, (b) wind speed simulated with default parameters (c) wind speed simulated with calibrated parameters, (d) wind speed bias between default and observations, and (e) wind speed bias between calibration and observations.}
    \label{spatial_ws10_calibration}
\end{figure}

\begin{figure}
    \centering
    \includegraphics[width=0.8\linewidth]{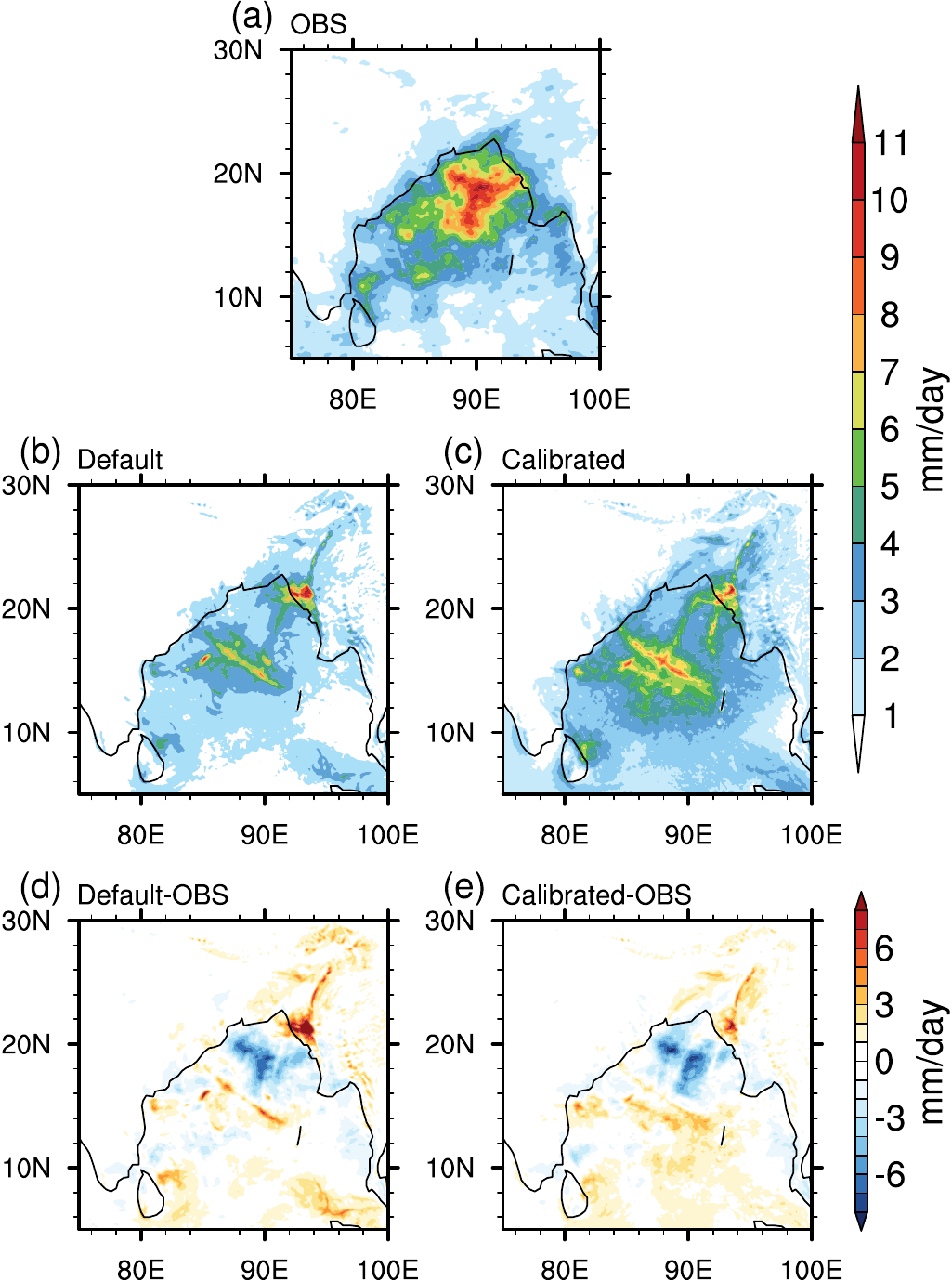}
    \caption{Spatial distributions of the domain averaged precipitation simulated with default and calibrated parameters, compared with the observations, for the calibration events (A-J). (a) observed precipitation, (b) precipitation simulated with default parameters (c) precipitation simulated with calibrated parameters, (d) precipitation bias between default and observations, and (e) precipitation bias between calibration and observations.}
    \label{spatial_precipitation_calibration}
\end{figure}


\begin{figure}
    \centering
    \includegraphics[width=0.6\textwidth,angle=90,origin=c]{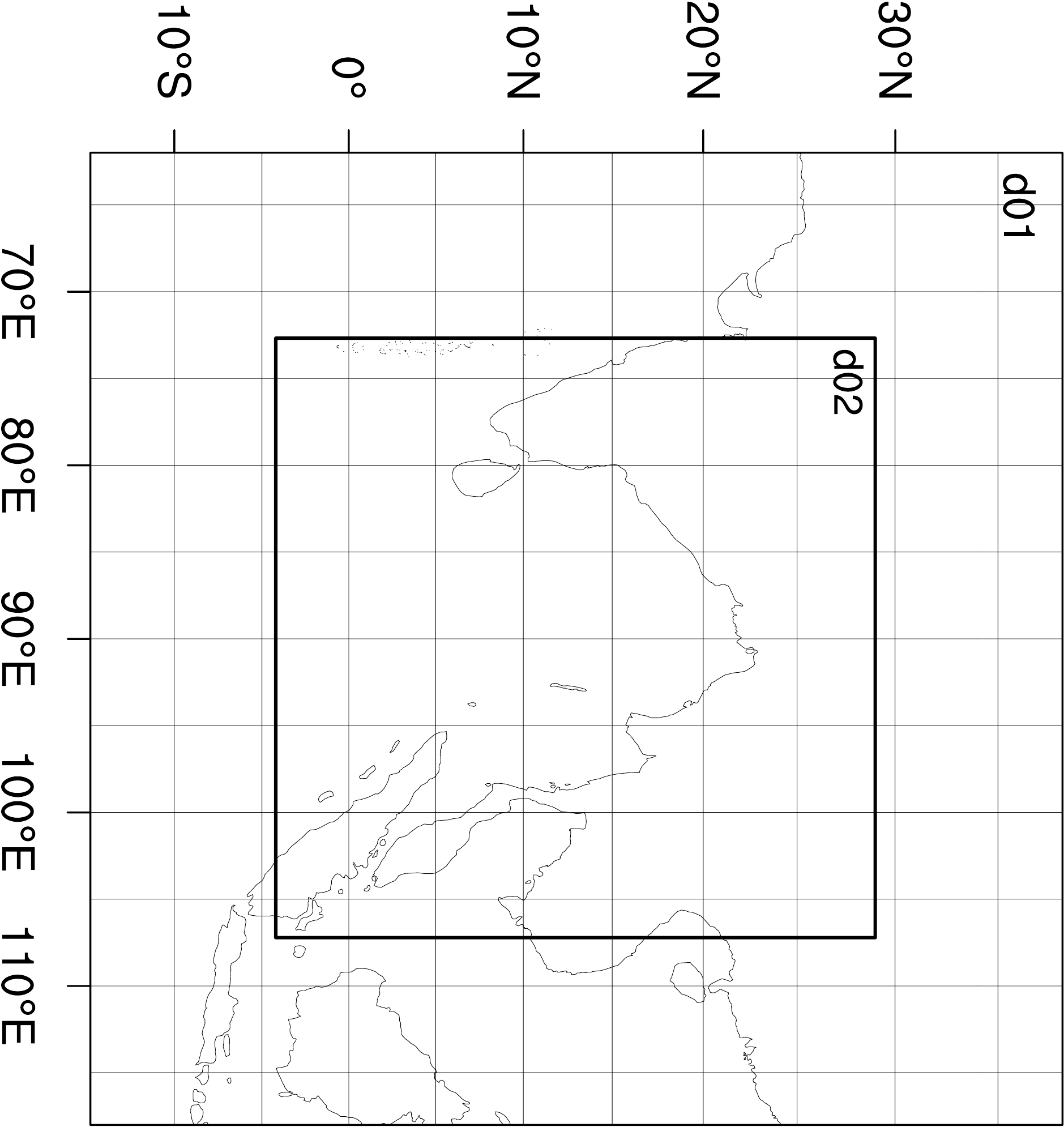}
    \caption{An illustration of the WRF model domain configuration. The parent domain (d01) consists \(520 \times 520\) grid points with 12km resolution and the nested domain (d02) consists of \(960 \times 960\) grid points with 4km resolution.}
    \label{domains_4km}
\end{figure}

